\documentclass[useAMS,usenatbib]{mnras}

\def\Zsol{\hbox{Z$_{\odot}$}}
\def\Msol{\hbox{M$_{\odot}$}}

\usepackage{epsfig,natbib,url,graphicx,float}
\usepackage{lscape,amsmath,amssymb}
\usepackage{booktabs,threeparttable}
\usepackage{longtable,deluxetable}
\usepackage{color}
\usepackage{soul} 
\usepackage{fancyvrb}
\usepackage{pifont}
\usepackage{ulem}
\usepackage{pdflscape}
\usepackage{caption}
\usepackage{subcaption}
\usepackage[dvipsnames]{xcolor} 

\newcommand{\hii}{H~{\sc ii}}
\newcommand{\hei}{He~{\sc i}}

\newcommand{\kms}{km\,s$^{-1}$}
\newcommand{\eld}{$N_{\rm e}$}

\newcommand{\elt}{$T_{\rm e}$}

\newcommand{\op}{O$^+$}

\newcommand{\opp}{O$^{2+}$}

\newcommand{\foiii}{[O~{\sc iii}]}

\newcommand{\foii}{[O~{\sc ii}]}
\newcommand{\fsii}{[S~{\sc ii}]}
\newcommand{\fsiii}{[S~{\sc iii}]}

\newcommand{\fnii}{[N~{\sc ii}]}

\newcommand{\hp}{H$^+$}

\newcommand{\ha}{H$\alpha$}
\newcommand{\hb}{H$\beta$}

\usepackage{caption}

\title[Exploring Chemical Homogeneity in Dwarf Galaxies]{Exploring Chemical Homogeneity in Dwarf Galaxies: A VLT-MUSE study of JKB~18}
\author[James et al. ]{Bethan L. James$^{1}$\thanks{E-mail:bjames@stsci.edu}, Nimisha Kumari$^{2}$, Andrew Emerick$^{3,4}$, Sergey E. Koposov$^{5,6}$, 
\newauthor Kristen B. W. McQuinn$^7$,
Daniel P. Stark$^{8}$, Vasily Belokurov$^6$,  
 \&  Roberto Maiolino$^2$\\
$^1$ AURA for ESA, Space Telescope Science Institute, 3700 San Martin Drive, Baltimore, MD 21218\\
$^2$ Kavli Institute for Cosmology, University of Cambridge, Madingley Road, Cambridge CB3 0HA, UK\\
$^3$ Carnegie Observatories, Pasadena, CA, 91101, USA \\
$^4$ TAPIR, California Institute of Technology, Pasadena, CA, 91125, USA \\
$^5$ McWilliams Center for Cosmology, Carnegie Mellon University, 5000 Forbes Ave, Pittsburgh, PA 15213, USA \\
$^{6}$ Institute of Astronomy, University of Cambridge, Madingley Road, Cambridge, CB3 0HA\\
$^7$ Rutgers University, Department of Physics and Astronomy, 136 Frelinghuysen Road, Piscataway, NJ 08854, USA\\
$^{8}$ Steward Observatory, The University of Arizona, 933 N Cherry Ave, Tucson, AZ, 85721, USA\\
\\
}

\begin{document}

\date{Accepted 2020 May 4. Received 2020 April 1; in original form 2019 November 8}

\pagerange{\pageref{firstpage}--\pageref{lastpage}} \pubyear{2020}

\maketitle

\label{firstpage}

\begin{abstract}
Deciphering the distribution of metals throughout galaxies is fundamental in our understanding of galaxy evolution. Nearby, low-metallicity, star-forming dwarf galaxies in particular can offer detailed insight into the metal-dependent processes that may have occurred within galaxies in the early Universe. Here we present VLT/MUSE observations of one such system, JKB~18, a blue diffuse dwarf galaxy with a metallicity of only 12+log(O/H)$=7.6\pm0.2$ ($\sim$0.08~\Zsol). Using high spatial-resolution integral field spectroscopy of the entire system, we calculate chemical abundances for individual \hii\ regions using the direct method and derive oxygen abundance maps using strong-line metallicity diagnostics. With large-scale dispersions in O/H, N/H and N/O of $\sim$0.5--0.6~dex and regions harbouring chemical abundances outside this 1$\sigma$ distribution, we deem JKB~18 to be chemically inhomogeneous. We explore this finding in the context of other chemically inhomogeneous dwarf galaxies and conclude that neither the accretion of metal-poor gas, short mixing timescales, or self-enrichment from Wolf-Rayet stars are accountable. Using a galaxy-scale, multi-phase, hydrodynamical simulation of a low-mass dwarf galaxy, we find that chemical inhomogeneities of this level may be attributable to the removal of gas via supernovae and the specific timing of the observations with respect to star-formation activity. This study not only draws attention to the fact that dwarf galaxies \textit{can} be chemically inhomogeneous, but also that the methods used in the assessment of this characteristic can be subject to bias.
\end{abstract}


\section{Introduction}

The `gas regulatory' or `bathtub' chemical processing frameworks of galaxy evolution, are thought to consist of four key components: star-formation,  chemical  enrichment,  outflows,  and accretion \citep[e.g.,][]{Lilly:2013,Dayal:2013, Peng:2014}. Each of these components is dependent on the other, operating together in the multi-phase interstellar medium via the exchange of energy and momentum. The metal content of the medium through which this exchange occurs can play a significant role - both in communicating the evolutionary \textit{history} of the galaxy and shaping the evolutionary \textit{future} of the galaxy. Via collisional excitation and recombination processes, metals provide efficient mechanisms through which gas can cool, condense, and eventually form stars \citep[e.g.,][]{Krumholz:2012}.  As these stars form and evolve, they eject their metals into the interstellar medium (ISM), via photospheric winds during the asymptotic giant branch (AGB) phase or through supernovae driven outflows, enriching the ISM and intergalactic medium (IGM) with chemo-evolutionary signatures that depend on the temperature and mass of the star. Since the winds of hot stars are driven by photon momentum transfer through metal line absorption, metals also drive the momentum and velocity of the wind \citep{Kudritzki:2000}. In turn, the metal content of accreted gas, combined with the metal content of the site of subsequently triggered star-formation, can provide important constraints on the enrichment history and structure of the IGM surrounding the galaxy \citep[][and references therein]{Putman:2017}.  As such, a deep and thorough understanding of the spatial distribution of metals in the gas throughout galaxies, is imperative in furthering our insight into galaxy evolution in general.

Given the important role of metals in deciphering global trends of galaxy evolution - such as the mass-metallicity relation \citep{Lequeux:1979,Tremonti:2004} and the fundamental metallicity relation \citep{Ellison:2008, Mannucci:2010} - much effort has been spent in investigating the spatial distribution of metals throughout star-forming galaxies. For several decades, metallicity gradients have been found to be prevalent in galactic discs, typically showing a negative relation as a function of galactic radius \citep[e.g.,][]{Searle:1971,Pagel:1981, Vila-Costas:1992,Zaritsky:1994,VanZee:1998,Moustakas:2010,Werk:2011} - which provided insight into processes that regulated the growth and assembly of galaxies such as inside-out growth and/or galaxy quenching. Inverted metallicity gradients have also been found at high redshifts \citep[e.g.,][]{Cresci:2010, Troncoso:2014}, which are thought to be due to an enhanced inflow of chemically pristine gas towards the center of the galaxy. Furthermore, metallicity gradients have been characterized at a higher level of detail during the past decade or so due to the advent of integral field units (IFUs). Thanks to large-scale surveys of spatially resolved spectroscopy, such as CALIFA \citep{Sanchez:2012}, MaNGA \citep{Bundy:2015}, and SAMI \citep{Croom:2012} we now know the shape of the gradient depends on the stellar mass of the galaxy \citep{Belfiore:2017,Goddard:2017, Poetrodjojo:2018}. While these works have provided significant insight into the role of metals in galaxy evolution, by concentrating almost entirely on massive ($M_{\star}>10^{10}$~\Msol), disc-dominated galaxies, they provide a limited view on the evolutionary processes involved in \textit{early} Universe. As discussed in \citet{Maiolino:2019}, measurements of radial gradients are deemed virtually inefficient at early epochs of galaxy evolution due to their large, irregular metallicity variations \citep[most likely due to more chaotic accretion and formation processes e.g.,][]{ForsterSchreiber:2018} and thus fail to represent the chemical complexity of such systems. 

Since current facilities cannot provide us with the detailed spectroscopic information required to investigate the spatial distribution of metals and chemical homogeneity of the first galaxies, we instead rely on nearby `analogues' of these complex systems. Nearby low-mass galaxies with extremely low metal content (i.e.,$<10$\%\ solar metallicity), currently or recently undergoing periods of intense star-formation provide close representations, while also enabling the spatial and spectral detail that we require for such studies. There are of course some caveats to this approach - for example, local extremely metal poor \citep[XMP, e.g.,][]{Papaderos:2008, Sanchez-Almeida:2016,Filho:2015} galaxies are known to have much lower specific star-formation rates \citep[0.1--1\,Gyr$^{-1}$,][]{James:2017} compared to star-forming systems at $z=$6--10 \citep[5--10\,Gyr$^{-1}$,][]{Lehnert:2015}, and thus different massive-star populations. Also, being in the local Universe, and thus having undergone previous periods of star-formation, such `analogues' cannot be true representatives of the first galaxies by design. Despite these set-backs, one such population of galaxies that we can use to our advantage are Blue Diffuse Dwarf galaxies (BDDs). BDD galaxies were brought to light by conducting a search on SDSS imaging data using the morphological properties of Leo P, one of the most metal-deficient dwarf galaxies known \citep[$\sim$1/34~\Zsol,][]{Skillman:2013}. The search uncovered $\sim$130 previously undetected low surface-brightness star-forming galaxies, a sub-sample of which were followed-up with MMT optical long-slit spectroscopy \citep{James:2015, James:2017}. Each of these galaxies are faint, blue systems, with isolated \hii\ regions embedded within a diffuse continuum. Optical long-slit spectroscopy confirms that $\sim$25\%\ of the sample were indeed extremely metal poor. Combined with SDSS imaging, optical spectra reveal them to be nearby systems (5--120~Mpc) with regions of ongoing and recent star-formation (i.e.,star-formation rates of $\sim$0.03--8$\times10^{-2}$~\Msol/yr, with average ages of $\sim$7 Myr) randomly distributed throughout their diffuse, chemically-unevolved gas. This combination of pristine environments in a pre- or post-starburst era offers an opportunity to explore and understand the interplay between star-formation and metallicity in `young' star-forming environments that we expect to see in the high-z Universe - in exceptional detail.

\begin{figure*}
\includegraphics[scale=0.45]{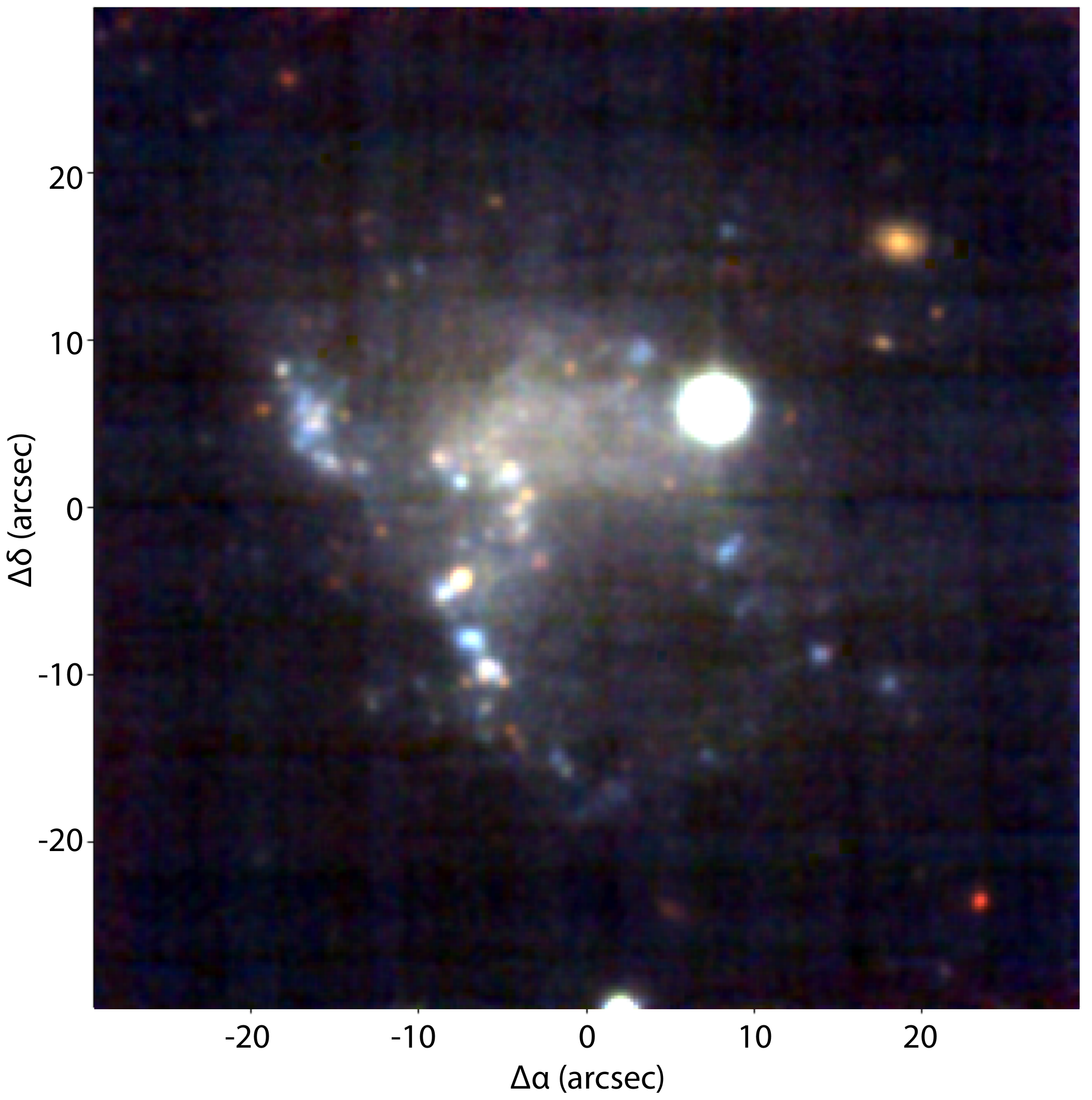}
\caption{Color composite image of JKB~18 ($\alpha=$09:21:27.173, $\delta=$07:21:50.70) derived from the VLT-MUSE data cube, created from 100~\AA\ channels at 4700~\AA\ (blue), 6050~\AA\ (green), and 8000~\AA\ (red). North is up, east is to the left.} 
\label{fig:color}
\end{figure*}

Here we present IFU observations of one such BDD, JKB~18 (Figure~\ref{fig:color}, Table~\ref{tab:gal}). As presented in \citet{James:2017}, this is a nearby ($z=0.004$), low metallicity (12+log(O/H)$=7.56\pm0.20$) system forming stars at an average rate of  $0.20\pm0.09\times10^{-2}$~\Msol/yr. We estimate a distance of $\sim$18~Mpc and a stellar mass of $\sim10^8$~\Msol, although due to radial velocities and peculiar motions, these estimates are subject to significant uncertainties. As shown in Figure~\ref{fig:color}, JKB~18 has an extremely interesting morphology, with numerous sites of ongoing star-formation scattered in a somewhat random fashion amongst a diffuse body. In this paper, we utilise the high sensitivity and high spatial resolution of VLT-MUSE observations to perform a detailed spatially-resolved spectroscopic study of this system, in order to investigate the distribution of metals throughout its ionized gas. Our goal is to explore the chemically inhomogeneous nature of its gas and understand it in the context of evolutionary processes such as star-formation, outflows, and gas accretion.

The paper is structured as follows: in Section~\ref{sec:obs}, we describe the VLT/MUSE observations and data reduction. Maps of flux, reddening, and emission line diagnostics for JKB~18 and the methods used to create them are described in Section~\ref{sec:lines}. In Section~\ref{sec:HIIregs} we explore the chemical homogeneity of the ionized gas, including direct-method oxygen abundances and nitrogen-to-oxygen ratios of individual \hii\ regions, and metallicity maps using strong line diagnostics. These findings are discussed in the context of other dwarf galaxies in Section~\ref{sec:disc}, and interpreted via a detailed galaxy-scale simulations of a similar, XMP system. In Section~\ref{sec:conc} we draw conclusions on our findings.

\begin{figure*}
\includegraphics[scale=0.6]{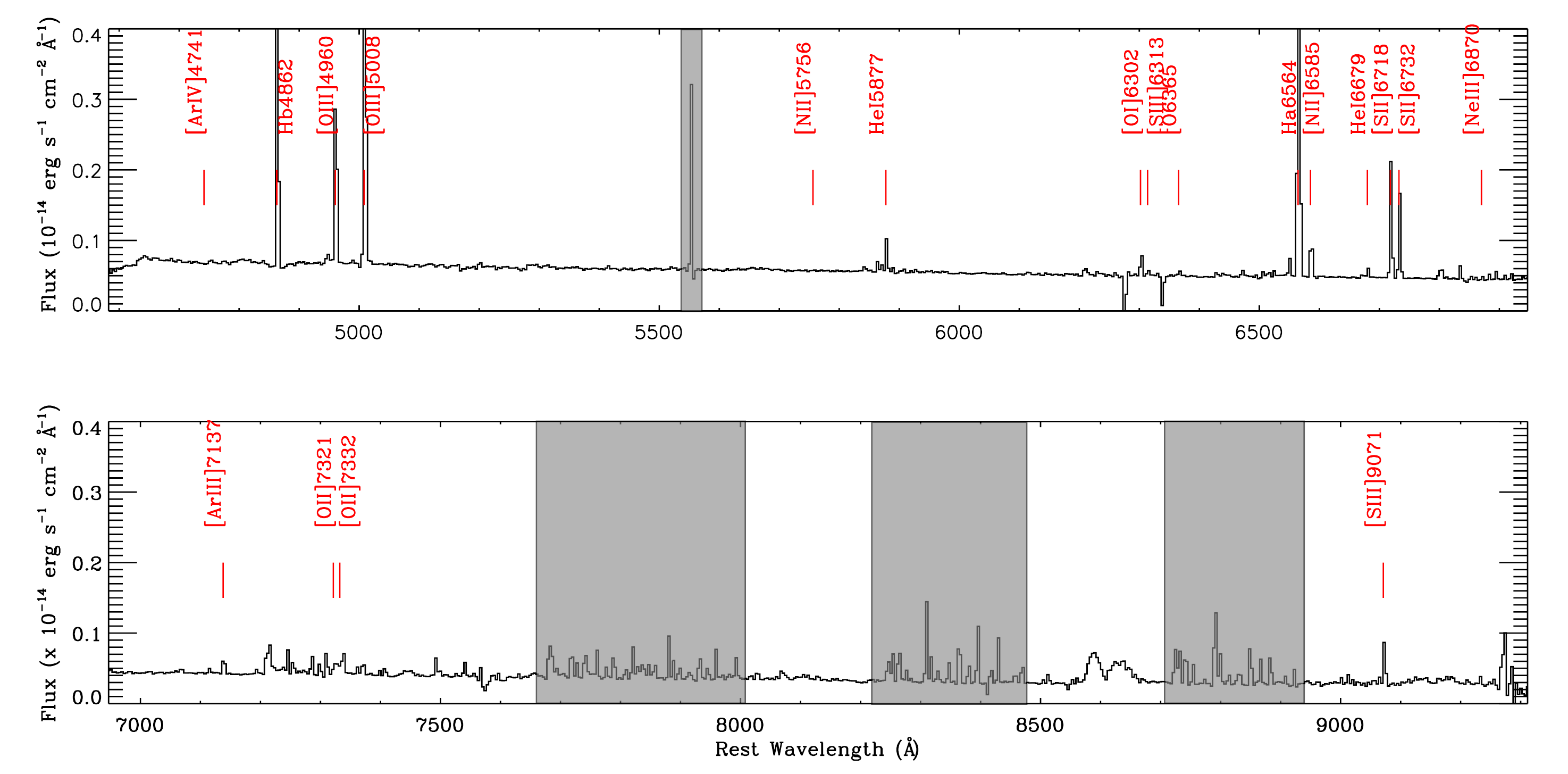}
\caption{Emission line spectrum summed over all 30 identified \hii\ regions throughout JKB~18 (as described in Section~\ref{sec:HIIregs}). Major emission lines are labelled in red. Grey rectangles correspond to regions of poor sky subtraction.} 
\label{fig:full_spec}
\end{figure*} 

\begin{table}
\caption{General properties of JKB~18, as presented in \citet{James:2017}}
\begin{center}
\begin{tabular}{|c|c|}
\hline
R.A. & 9:21:27.173 \\
Dec. & 07:21:50.70 \\
$z$ & 0.004\\
Distance & $\sim$18~Mpc\\
12+log(O/H) & 7.56$\pm$0.20\\
Mass & $\sim10^8$~\Msol \\
\hline
\end{tabular}
\end{center}
\label{tab:gal}
\end{table}%

\section{Observations \&\ Data Reduction}\label{sec:obs}

Observations of JKB~18 (Table~\ref{tab:gal}) were made using VLT-MUSE on 18/11/2015 and 10/1/2016 under program 096.B-0212(A) (PI: James). Data were taken in wide-field-mode, with extended wavelength coverage (4650-9300~\AA) and a resolution of $R\sim$1770--3950. The $1\times1$ arcminute field-of-view of MUSE fully covers JKB~18, providing 0.2\arcsec $\times$ 0.2\arcsec plate-scale spatial resolution for each of the 300$\times$300 spaxels (providing 90,000 spectra in total). Observing blocks (OBs) were broken down into $5\times275$~s and $5\times600$~s exposures, for a total of 1.2~hours integration time on source. Separate sky observations were taken for each OB and the seeing was reported to be 0.9\arcsec\ ($\sim$78~pc at the distance of JKB~18) for both sets of observations, as such it was necessary to bin our datacube using a two-dimensional 3~pixel boxcar kernel before performing emission line fitting to avoid oversampling.

Although the MUSE data cubes had already undergone data reduction via the ESO-MUSE pipeline, upon inspection the cubes were found to have poorly subtracted sky lines. As such, the cubes were re-reduced using the MUSE pipeline version 2.4.2\footnote{\url{ https://www.eso.org/sci/software/pipelines/muse/}}, with stricter constraints on the sky subtraction parameters (such as a non-modified sky subtraction to mitigate contamination of the continuum from the sky spectrum, and for the sky frames a usable pixel fraction of 40\%\ and an ignore pixel fraction of 10\%). Specifically, we implemented this using {\sc MUSEpack}, a python-based wrapper for ESO's MUSE pipeline as described in \citet{Zeidler:2019}. The pipeline covers flat-field removal, wavelength calibration, flux calibration, and sky removal, finally combining the cubes from each OB into a single datacube. The wavelength solution and line-spread function (LSF) of the MUSE instrument is very stable. Nevertheless, we used a 90 degrees dither pattern strategy, as suggested by the MUSE handbook, to not only remove cosmic rays and detector defects, but also to reduce possible small offsets in the wavelength and flux calibration. Furthermore, the stability of the wavelength and LSF was checked by fitting multiple sky lines across several sky-only regions across the cube. Variation in both parameters were found to be negligible.

\section{Emission Line Maps}\label{sec:lines}
The full MUSE spectrum, integrated across JKB~18 can be seen in Figure~\ref{fig:full_spec}.  All of the strong emission lines typical of \hii\ regions can be seen, along with the \hei\ series, the \fsiii~$\lambda$6312 auroral line, the \foii~$\lambda\lambda$7321, 7332 doublet, and \fsiii~$\lambda$9071 (the companion of this doublet line, \fsiii~$\lambda$9531, is not covered by the MUSE wavelength range). The methodology used to fit each of these emission lines throughout JKB~18 and the resultant emission line maps are described below.

\subsection{Automated Line Fitting}
For each spaxel, we performed emission line fitting using {\sc ALFA} (Automated Line Fitting Algorithm)\footnote{https://www.nebulousresearch.org/codes/alfa/}, an automated line fitting tool optimised to work on MUSE data cubes. A full description of ALFA can be found in \citet{Wesson:2016}. The tool's novel approach consists of constructing a synthetic spectrum to match the observations, derived from a line catalogue input from the user, then optimising the parameters of all the Gaussian line profiles by means of a generic algorithm. In addition to the Gaussian parameters and uncertainties for each line (estimated using the noise structure of the residuals), the tool additionally outputs the fitted spectrum and the fit to the continuum. It should be noted that stellar absorption features were not seen in individual spaxel spectra. In order to estimate the effects of any underlying stellar absorption, tests were performed on the highest S/N single-spaxel spectra by additionally including Balmer-line absorption features while fitting. Accounting for the absorption had a negligible affect on the final parameters of the emission line fits.

\subsection{Structure of the Ionized Gas}
Figure~\ref{fig:Halpha} (top panel) shows the structure of the \ha\ emitting gas throughout JKB~18. The emission appears to be clumpy, with a multitude of \hii\ regions ($\sim$100--130~pc in size, as found by the \hii\ region fitting algorithm described in Section~\ref{sec:HIIregs} and resolved at the resolution of our observations) distributed randomly across the galaxy. This morphology is in contrast to blue compact dwarf (BCD) galaxies which typically have a compact, central emitting region. The structure suggests little or no large-scale order, although some areas do appear to be connected as streams or arms of emission.  A detailed analysis of each \hii\ region is described in Section~\ref{sec:HIIregs}.

Also shown in Figure~\ref{fig:Halpha} are the radial velocity and velocity dispersion maps for \ha. The radial velocity map has been corrected for the systemic velocity of 1378~\kms, as derived from the redshift of the \ha\ emission line in the integrated spectrum, and the velocity dispersion map has been corrected for the instrumental resolution at \ha\ of $\sim$2~\AA\ ($\sim$88~\kms\, as measured from a selection of sky lines).  We can see from the radial velocity map that there is indeed some large-scale gas structure throughout JKB~18, in that the gas appears to be rotating as a large body. The range in velocities across the whole system is relatively small, on the order of $\pm20-30$~\kms\ (although given the unknown inclination of the system, this rotation could potentially be larger).  Within this radial velocity frame, there is one region that appears to be disconnected from the galaxy where there is a rapid jump to $\sim - 20$~\kms in the most south-western clump of emission. The upper north-eastern `arm' of emission ($x=\sim-15$ arcsec,$y=\sim10$ arcsec) also appear to be disjoint from the rest of the galaxy in that all the gas is moving towards us at a constant 30~\kms.  Conversely, the velocity dispersion map shows a surprising lack of structure. The large majority of the gas has a dispersion between 100 and 120~\kms. This suggests that there is little or no movement within the gas due to e.g., large-scale in/outflowing gas.

\begin{center}
\includegraphics[scale=0.5]{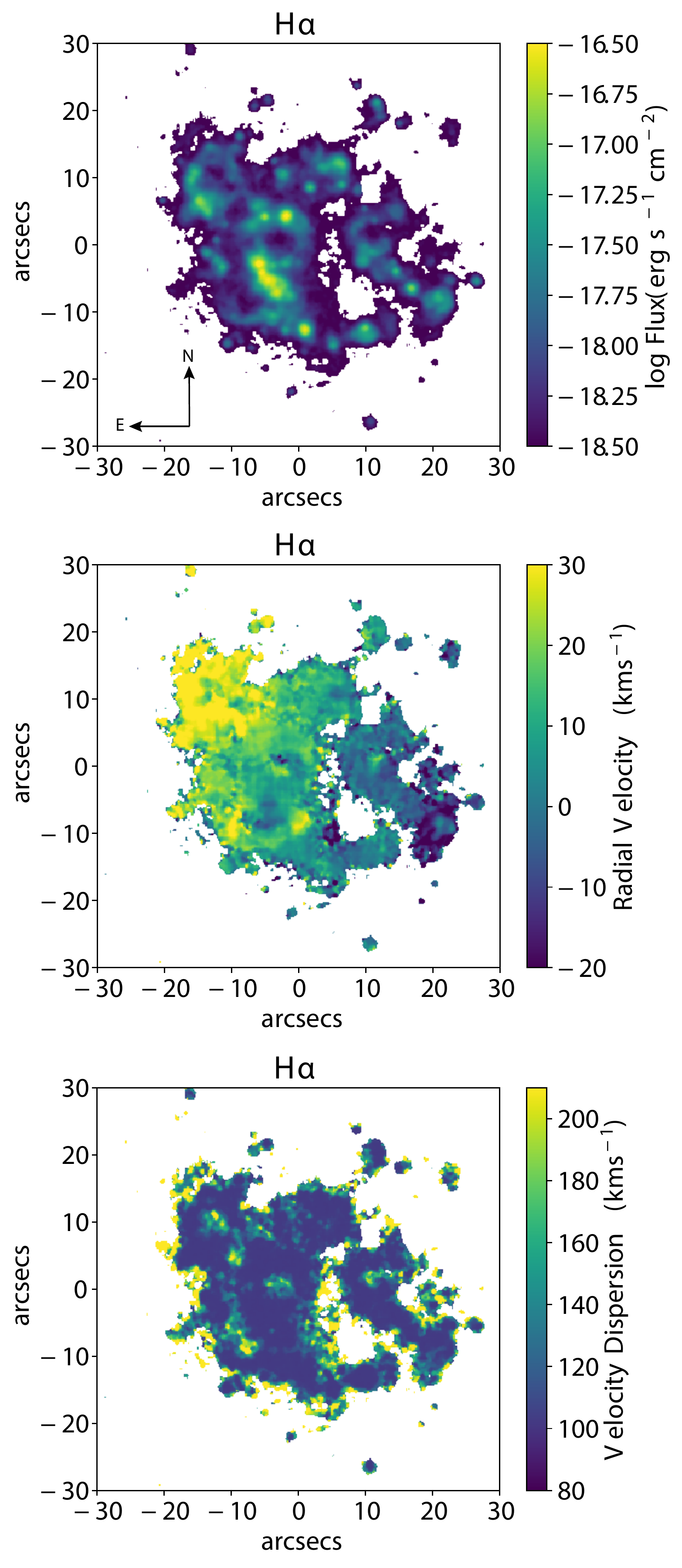}
\captionof{figure}{Maps of flux, radial velocity, and velocity dispersion (FWHM) of JKB~18, as measured from the \ha\ emission line across the MUSE FoV. White spaxels correspond to those with $<3\sigma$ detections.} 
\label{fig:Halpha}
\end{center} 

\begin{figure*}
\includegraphics[scale=0.5]{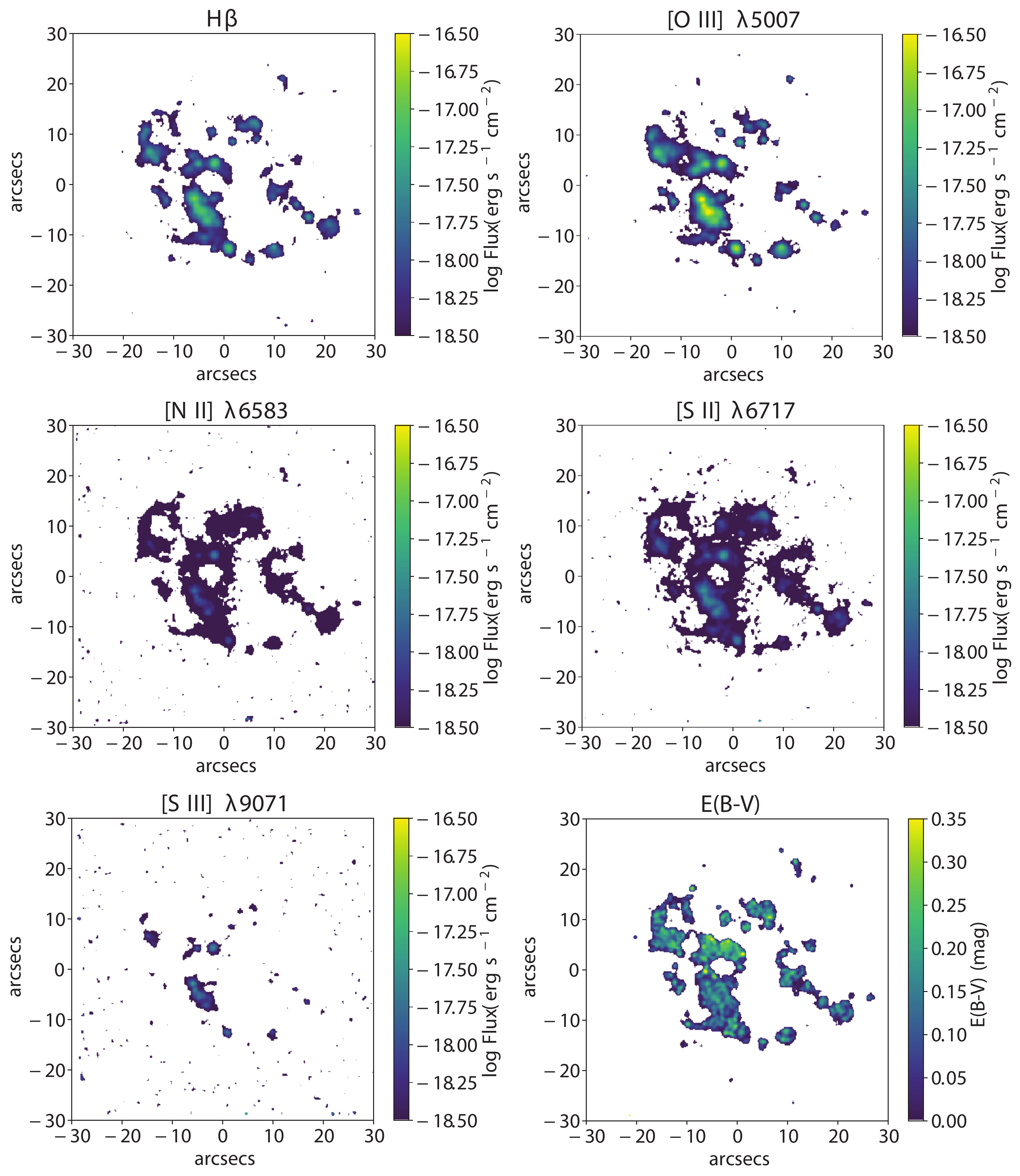}
\caption{Emission line maps of \hb, \foiii~$\lambda$5007, \fnii~$\lambda$6583, \fsii~$\lambda$6716, and \fsiii~$\lambda$9071. The bottom right-hand panel shows the map of E(B-V) throughout JKB~18, used to correct emission line maps for reddening when necessary. North is up and east is to the left. White spaxels correspond to those with $<3\sigma$ detections.} 
\label{fig:maps}
\end{figure*} 

\begin{figure}
\includegraphics[scale=0.55]{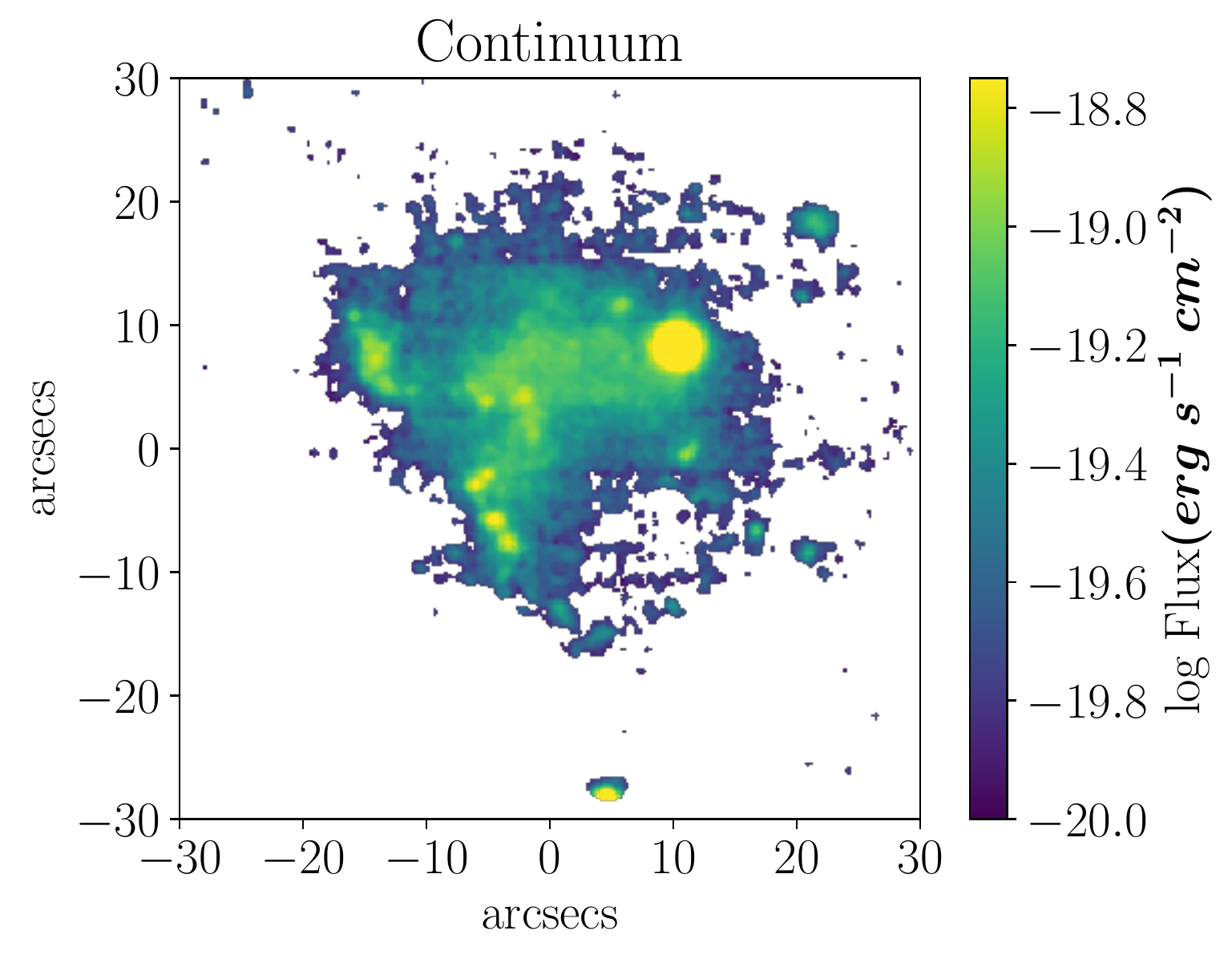}
\caption{Continuum flux map, derived from emission line free wavelength range between 4910--4960~\AA. North is up, east is to the left. White spaxels correspond to those with $<10\%$ uncertainty in the continuum flux.}
\label{fig:cont_map}
\end{figure} 

The remaining `strong' emission lines - \hb, \foiii~$\lambda$5007, and \fsii~$\lambda$6716 - were mapped on a spaxel by spaxel basis throughout JKB~18 and the resultant maps are shown in Figure~\ref{fig:maps}. Here the emission can be seen throughout all the main star-forming regions, following closely the \ha\ morphology shown in Figure~\ref{fig:Halpha}. We have also mapped the somewhat weaker lines,  \fnii~$\lambda6584$ and  \fsiii~$\lambda$9071. For the latter, the emission is detected in only a few of the star-forming knots.

In Figure~\ref{fig:cont_map} we show the continuum flux map summed over 4910--4960~\AA. In general, the continuum emission from the stellar populations follows that of the emission from the ionized gas. However, there are instances where the \hii\ region emission is not accompanied by continuum emission, such as those located in the south-west of the system. It should be noted that the bright continuum source at $x\sim10'',y\sim10''$ is a foreground star.

\subsection{E(B-V) map}

The extinction throughout JKB~18 was calculated using the $F$(\ha)/$F$(\hb) emission line map ratio, adopting the case B recombination ratio (\ha/\hb=2.79, appropriate for gas with \elt$=14,000$~K and \eld=40~cm$^{-3}$, the average electron temperature and electron density, respectively, calculated from the \hii\ regions in JKB~18, Section~\ref{sec:HIIregs}) and the Large Magellanic Cloud extinction curve \citep{Fitzpatrick:1999}. The resultant $E(B-V)$ map can be seen in the bottom right-hand panel of Figure~\ref{fig:maps}, with values ranging from $E(B-V)=$ 0.0 to 0.3~mag. The areas of highest extinction roughly follow the peaks in the \ha\ flux distribution (Fig.~\ref{fig:Halpha}, top-panel).

\subsection{Sources of Ionization across JKB~18}
\begin{figure*}
\includegraphics[scale=0.4]{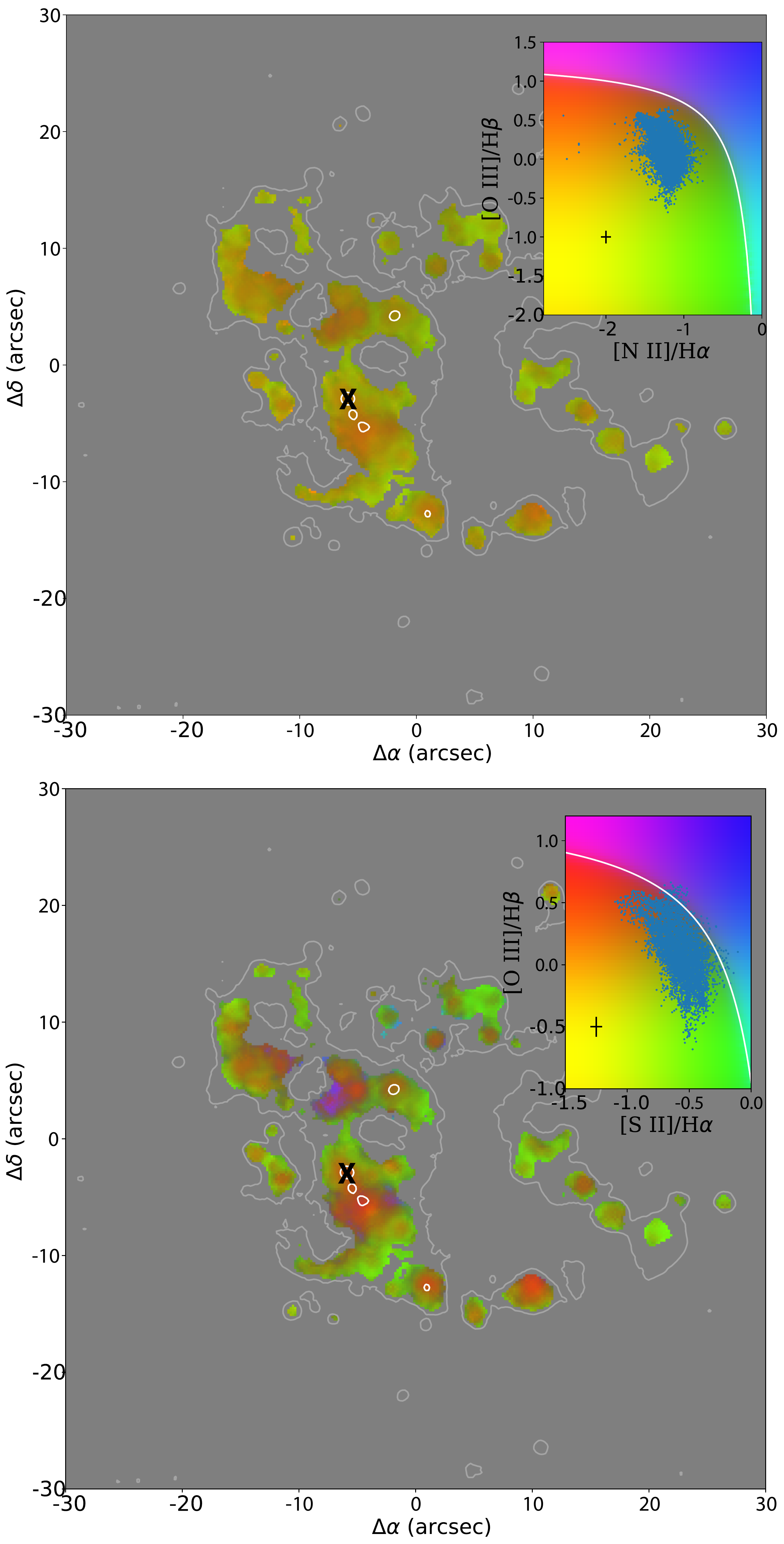}
\caption{Emission line diagnostic maps of JKB~18 showing log(\fnii/\ha) vs. log(\foiii/\hb) (top-panel) and log(\fsii/\ha) vs. log(\foiii/\hb) (bottom-panel) derived from the line ratio images in Figure~\ref{fig:maps} and color-coded according to their location within the BPT diagram, shown in the inset of each panel. Overlaid contours represent the morphology of \ha\ emission (Figure~\ref{fig:Halpha}) and 'X' denotes the central star-forming region. The “maximum starburst line” from \citet{Kewley:2001} is overlaid in white. The black cross overlaid on each diagnostic diagram represents the average error for the emission line diagnostic ratio. Grey spaxels correspond to those with $<3\sigma$ detections. North is up, east is to the left.} 
\label{fig:BPT}
\end{figure*}

In order to explore the sources of ionisation within JKB~18, in Figure~\ref{fig:BPT} we show emission line diagnostic images of the galaxy. Here, each spaxel is colour-coded according to its placement within the \fnii/\ha\ or \fsii/\ha\ versus \foiii/\hb\ diagram. These emission line diagnostic diagrams or `BPT' diagrams separate emission line ratios according to the source of ionising radiation, with the overlaid `maximum starburst line' from \citet{Kewley:2001} separating photoionised gas (below the line) from harder ionising sources such as shocks and active galactic nuclei (above the line). Since all of the emission line ratios from JKB~18 lie below the solid line, we can confirm that the gas is dominated by photoionization, however, as demonstrated by \citet{Plat:2019} and \citet{Kewley:2013}, given the low metallicity of the gas, contributions from shock excitation cannot be ruled out. Both diagnostics maps show an interesting ionization structure - while the central star-forming region (marked as `X') shows (an expected) high \foiii/\hb, other regions of similar excitation lie at the \textit{edges} of some star-forming regions and are offset from the areas of strongest star-formation (as indicated by the overlaid \ha\ contours), suggesting that the ionizing photons are not propagating in a uniform direction the center of the ionized gas and instead preferentially move towards the outer boundary of the ionized gas. There are clear gradients in decreasing \foiii/\hb\ as you move away from the edges of star-forming regions, an effect even more prominent in the \fsii/\ha\ vs. \foiii/\hb\ map. In particular, there is a region located at x=$-5$~arcsec, y=$+5$~arcsec which shows particularly high \foiii/\hb\ ratio for its \fsii/\ha\ ratio, very close to a gap in the ionised gas, with similar high values $\sim$3~arcsec eastward. This region, along with other edge-enhanced areas could signify shock-excited gas due to interactions with the surrounding ISM.  Overall these detailed diagnostic maps reveal a complex and inhomogeneous ionisation structure with widely varying levels of radiation field hardness throughout.

\begin{figure*}
\includegraphics[width=1\textwidth]{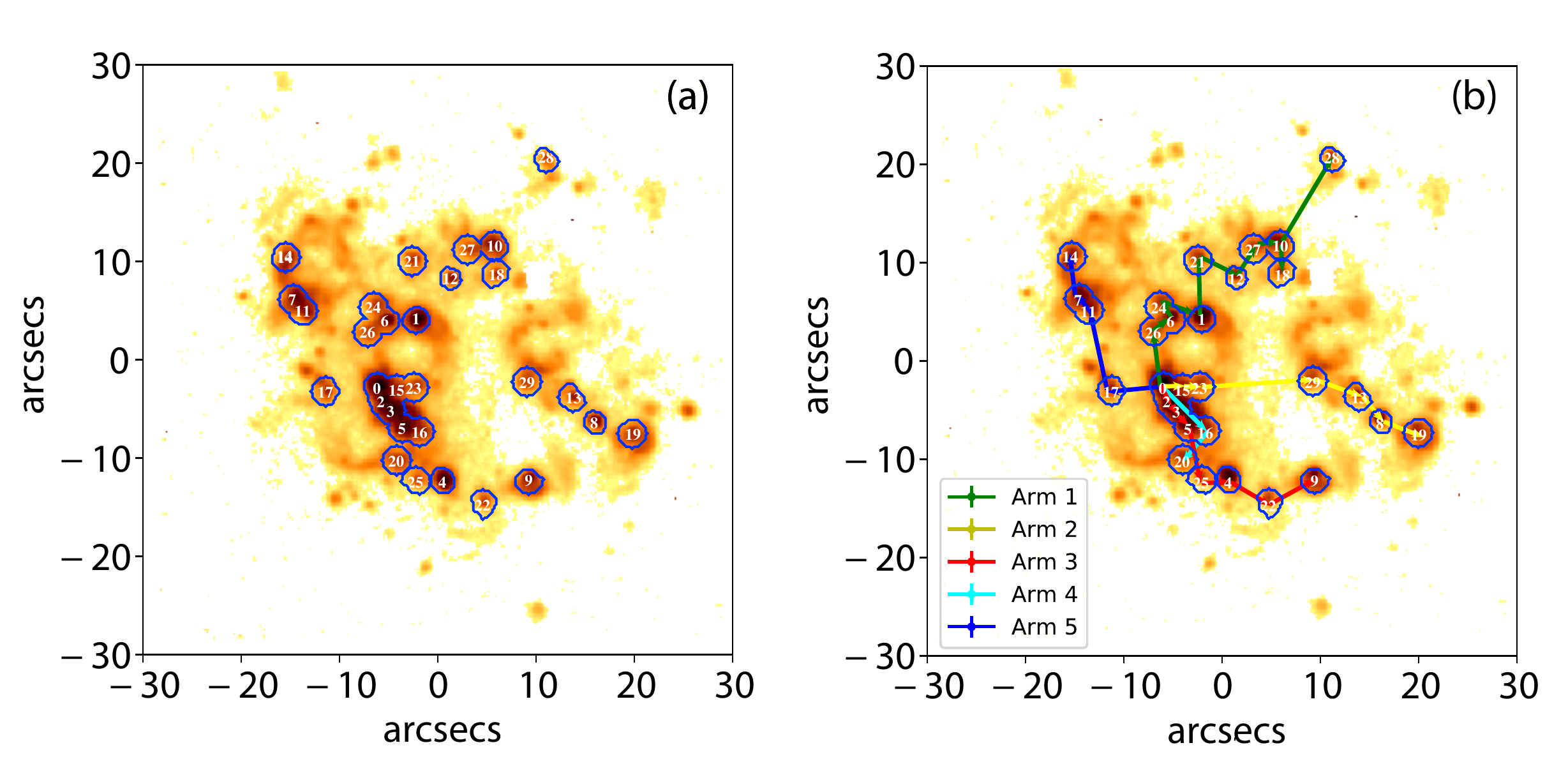}
\caption{(\textit{a}): Individual \hii\ regions identified using the \hii\ region binning algorithm described in Section~\ref{sec:HIIregs} overlaid on the \ha\ emission line map (Fig.~\ref{fig:Halpha}, upper panel). Each of the 30 \hii\ regions are labeled, with the blue outline denoting the `edge' of the \hii\ region, within which all spaxels are summed for the individual \hii\ region spectra. (\textit{b}): individual \hii\ regions with `arm' like structures overlaid. Structures were identified using the \ha\ flux, radial velocity, and FWHM maps shown in Figure~\ref{fig:Halpha} (as discussed in Section~\ref{sec:HIIregs}). North is up and east is to the left.} 
\label{fig:HIIreg}
\end{figure*}

\begin{table*}
\begin{center}
\begin{footnotesize}
\begin{tabular}{|c|ccccccccc|}
\hline
\hii\ Reg ID & \eld(\fsii ) & \elt(\fsiii)  & \elt(\foii) & \op/\hp & \opp /\hp  & 12+log(O/H) & log(N/O) & 12+log(N/H) & [O/H] \\
 & (cm$^{-3}$)  & (K) & (K) & ($\times10^{-5}$) & ($\times10^{-5}$) & Z[Direct] & &  & Z[O3N2] \\ 
\hline
0& 
13$\pm$7.8 & 
15300$\pm$490&
20100$\pm$586&
   0.70$\pm$   0.08&
   2.20$\pm$   0.18&
   7.46$\pm$   0.03&
  -1.15$\pm$   0.02&
   6.31$\pm$   0.02&
8.33$\pm$0.05 \\
1& 
41$\pm$9.5 & 
14200$\pm$619&
18600$\pm$740&
   1.15$\pm$   0.18&
   2.10$\pm$   0.24&
   7.51$\pm$   0.04&
  -1.07$\pm$   0.03&
   6.44$\pm$   0.03&
8.43$\pm$0.04 \\
2& 
22$\pm$11 & 
12800$\pm$605&
16600$\pm$723&
   1.26$\pm$   0.24&
   4.78$\pm$   0.65&
   7.78$\pm$   0.05&
  -1.28$\pm$   0.04&
   6.50$\pm$   0.04&
8.27$\pm$0.05 \\
3& 
28$\pm$13 & 
12400$\pm$668&
15900$\pm$798&
   1.02$\pm$   0.25&
   5.65$\pm$   0.91&
   7.82$\pm$   0.06&
  -1.27$\pm$   0.04&
   6.56$\pm$   0.04&
8.22$\pm$0.04 \\
4& 
36$\pm$14 & 
14400$\pm$947&
18800$\pm$1130&
   1.03$\pm$   0.24&
   2.59$\pm$   0.46&
   7.56$\pm$   0.06&
  -1.28$\pm$   0.04&
   6.28$\pm$   0.04&
8.34$\pm$0.06 \\
5& 
13$\pm$8.8 & 
11300$\pm$823&
14400$\pm$984&
   2.69$\pm$   0.88&
   4.32$\pm$   1.02&
   7.85$\pm$   0.08&
  -1.43$\pm$   0.06&
   6.41$\pm$   0.06&
8.36$\pm$0.06 \\
6& 
36$\pm$19 & 
15100$\pm$1500&
19800$\pm$1790&
   0.70$\pm$   0.25&
   3.04$\pm$   0.79&
   7.57$\pm$   0.10&
  -1.01$\pm$   0.07&
   6.56$\pm$   0.07&
8.34$\pm$0.05 \\
7& 
18$\pm$12 & 
16600$\pm$1480&
22000$\pm$1770&
   0.61$\pm$   0.17&
   1.51$\pm$   0.32&
   7.33$\pm$   0.08&
  -1.12$\pm$   0.05&
   6.21$\pm$   0.05&
8.37$\pm$0.04 \\
8& 
12$\pm$12 & 
$<$11700&
$<$15000&
$>$   2.14&
$>$   2.44&
$>$   7.66&
--&
$>$   6.39&
8.47$\pm$0.05 \\
9& 
18$^{+18}_{-18}$ & 
$<$9580&
$<$12000&
$>$   3.64&
$>$  11.97&
$>$   8.19&
--&
$>$   6.66&
8.29$\pm$0.07 \\
10& 
17$\pm$10. & 
$<$12500&
$<$16200&
$>$   1.49&
$>$   1.08&
$>$   7.41&
--&
$>$   6.29&
8.53$\pm$0.03 \\
11& 
41$\pm$25 & 
16600$\pm$1480$^\star$&
22000$\pm$1770&
   0.63$\pm$   0.20&
   1.16$\pm$   0.25&
   7.25$\pm$   0.08&
  -1.12$\pm$   0.06&
   6.13$\pm$   0.06&
8.41$\pm$0.04 \\
12& 
1.7$^{+4.3}_{-1.7}$ & 
$<$11600&
$<$14900&
$>$   3.47&
$>$   4.78&
$>$   7.92&
--&
$>$   6.59&
8.44$\pm$0.06 \\
13& 
100$\pm$44 & 
$<$13400&
$<$17500&
$>$   0.99&
$>$   2.53&
$>$   7.55&
--&
$>$   6.38&
8.41$\pm$0.07 \\
14& 
86$\pm$27 & 
$<$12600&
$<$16200&
$>$   1.50&
$>$   1.82&
$>$   7.52&
--&
$>$   6.29&
8.45$\pm$0.06 \\
15& 
9.5$^{+10.}_{-9.5}$ & 
15300$\pm$514$^\star$&
20100$\pm$614&
   0.75$\pm$   0.19&
   1.31$\pm$   0.12&
   7.31$\pm$   0.05&
  -1.01$\pm$   0.03&
   6.31$\pm$   0.03&
8.43$\pm$0.05 \\
16& 
14$\pm$13 & 
$<$10500&
$<$13300&
$>$   2.21&
$>$   3.42&
$>$   7.75&
--&
$>$   6.53&
8.45$\pm$0.04 \\
17& 
40$\pm$29 & 
$<$15800&
$<$20900&
$>$   0.69&
$>$   1.27&
$>$   7.29&
--&
$>$   6.02&
8.40$\pm$0.08 \\
18& 
20$\pm$19 & 
$<$15400&
$<$20300&
$>$   0.59&
$>$   1.57&
$>$   7.34&
--&
$>$   6.48&
8.45$\pm$0.05 \\
19& 
43$\pm$21 & 
$<$19900&
$<$26700&
$>$   0.19&
$>$   0.20&
$>$   6.59&
--&
$>$   5.98&
8.56$\pm$0.04 \\
20& 
24$\pm$19 & 
$<$12400&
$<$16000&
$>$   2.25&
$>$   1.38&
$>$   7.56&
--&
$>$   6.19&
8.50$\pm$0.04 \\
21& 
33$\pm$20 & 
$<$14700&
$<$19300&
$>$   0.79&
$>$   1.27&
$>$   7.31&
--&
$>$   6.44&
8.49$\pm$0.04 \\
22& 
180$\pm$71 & 
$<$17800&
$<$23700&
$>$   0.24&
$>$   0.99&
$>$   7.09&
--&
$>$   6.33&
8.43$\pm$0.05 \\
23& 
59$\pm$36 & 
$<$15300&
$<$20100&
$>$   0.62&
$>$   1.14&
$>$   7.25&
--&
$>$   6.31&
8.47$\pm$0.04 \\
24& 
32$\pm$27 & 
$<$11700&
$<$15000&
$>$   1.43&
$>$   4.79&
$>$   7.79&
--&
$>$   6.69&
8.36$\pm$0.04 \\
25& 
23$\pm$21 & 
$<$25000&
$<$33900&
$>$   0.26&
$>$   0.29&
$>$   6.74&
--&
$>$   5.76&
8.49$\pm$0.04 \\
26& 
180$\pm$39 & 
15100$\pm$1540$^\star$&
19900$\pm$1840&
   0.83$\pm$   0.34&
   3.50$\pm$   0.94&
   7.64$\pm$   0.10&
  -1.09$\pm$   0.07&
   6.54$\pm$   0.07&
8.35$\pm$0.07 \\
27& 
23$\pm$18 & 
$<$11800&
$<$15100&
$>$   1.40&
$>$   2.24&
$>$   7.56&
--&
$>$   6.62&
8.50$\pm$0.04 \\
28& 
48$\pm$44 & 
$<$16300&
$<$21500&
$>$   0.93&
$>$   0.88&
$>$   7.26&
--&
$>$   5.96&
8.42$\pm$0.08 \\
29& 
17$\pm$16 & 
$<$16800&
$<$22200&
$>$   0.16&
$>$   0.48&
$>$   6.81&
--&
$>$   6.43&
8.29$\pm$0.07 \\
\hline
\end{tabular}
\label{tab:abunds}
\end{footnotesize}
\end{center}

\caption{Ionic and elemental abundances for individual \hii\ regions throughout JKB~18, derived from the emission line measurements integrated over each \hii\ region and discussed in Section~\ref{sec:HIIregs}. Electron density and temperatures are also listed. Electron temperatures marked with `$^*$' are those for which the neighbouring \hii\ region temperature was adopted as a result of unfeasibly high electron temperatures being measured, despite $>3\sigma$ detections on the \fsiii~$\lambda$6312 auroral line. Z[O3N2] values correspond to \hii\ region averages and standard deviations, as measured from the Z[O3N2] map shown in Figure~\ref{fig:O3N2map}.}\label{tab:abunds} 
\end{table*}

\section{Chemical Abundances} \label{sec:HIIregs}
\subsection{Direct method abundances of Individual \hii\ regions}
 Since the auroral line \fsiii~$\lambda$6312 was not detected at a 3-sigma level on a spaxel-by-spaxel basis, it was not possible to derive a `direct method' abundance map. Instead, we binned the spectra across each of the \hii\ regions throughout the galaxy using a publicly-available \textsc{python}-based package \textsc{ifuanal}\footnote{https://ifuanal.readthedocs.io/en/latest/} and derive direct method chemical abundances from the spectra integrated over each \hii\ region. While this reduces our spatial resolution, the approach is deemed sufficient since temperatures derived from integrating over individual \hii\ regions are good representations of the spatially resolved average value \citep[e.g.,][]{Kumari:2019a}.

In order to identify and extract spectra of individual \hii\ regions, we used the \textsc{ifuanal}-implementation of an \hii\ region binning algorithm that is based on the \textsc{hiiexplorer}\footnote{http://www.caha.es/sanchez/HII\_explorer/} algorithm \citep{Sanchez:2012, Galbany:2016}. Briefly, this algorithm creates a two-dimensional continuum-subtracted flux map of H$\alpha$ emission line and identifies emission peaks with flux values equal to or above a specified minimum peak flux value in the H$\alpha$ line map. Starting with the brightest one, these emission peaks act as seeds for creating the bins which include all pixels within a specified maximum radius, with flux values above a minimum threshold and at least 10\% of the seed pixel flux but exclude any pixel which is already included in a previous bin. An optimum bin size of seven pixels and minimum threshold flux of 10$^{-19}$ erg s$^{-1}$ cm$^{-2}$ was found to maximize the number of \hii\ regions in JKB~18, while also preserving a sufficient S/N ($\gtrsim8$) in each spectrum for the analysis performed here.

Figure~\ref{fig:HIIreg} shows the location of the 30 individual \hii\ regions identified using the above method, where their IDs (0--29) are marked in the decreasing order of the flux of the emission peak/seed. The emission line fluxes for each \hii\ region are provided as online material.  In an attempt to discern some morphological structure to JKB~18, in the right-hand panel of Figure~\ref{fig:HIIreg} we overlay a series of `arm' like structures on the individual \hii\ regions. These structures were established using the \ha\ emission, radial velocity and FWHM maps (Fig.~\ref{fig:Halpha}) such that the ionised gas in the `arms' have similar kinematic properties (i.e., radial velocities and dispersions), and therefore may be connected to one another. We affirm here that these rudimentary structures are not binding in that they may or may not relate accurately to the actual structure or form of JKB~18, especially when one considers projection effects, and are merely used here to guide the analysis in the form of radial plots presented in the following sections.

\begin{figure*}
\includegraphics[width=1\textwidth]{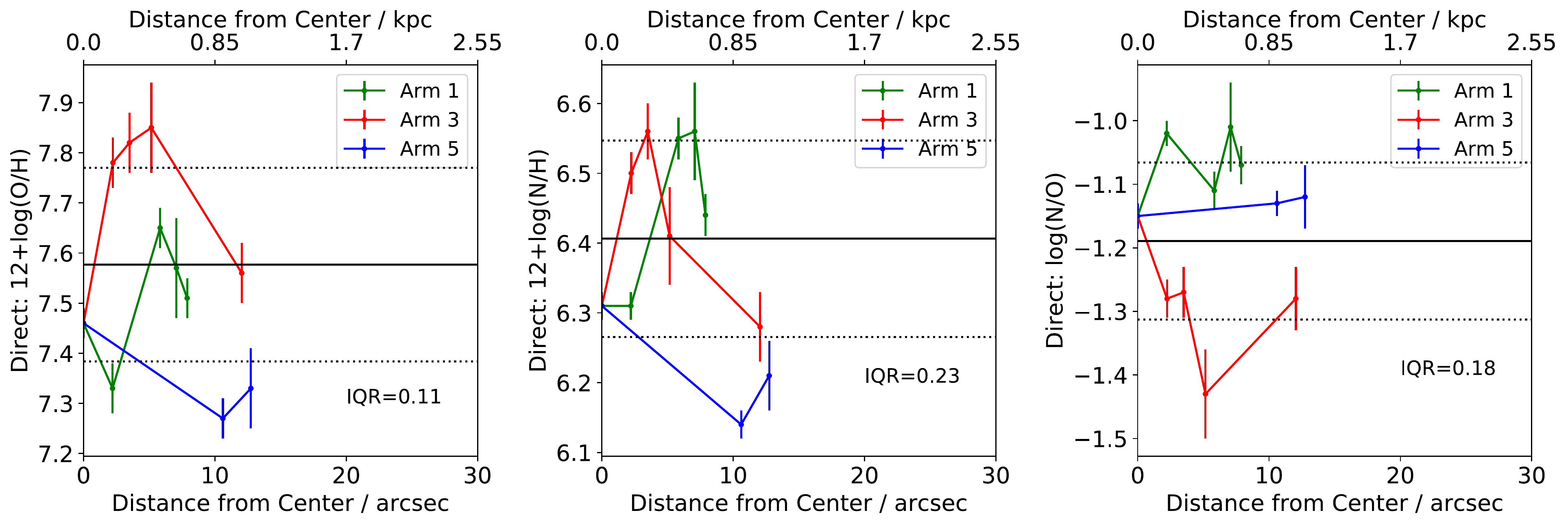}
\captionof{figure}{Direct method abundance measurements for individual \hii\ regions (shown in Figure~\ref{fig:HIIreg} and listed in Table~\ref{tab:abunds}) as a function of distance from the central \hii\ region in JKB~18 for O/H (left-panel), N/H (middle-panel) and N/O (right panel). Here we show only those \hii\ regions whose electron temperature was measured with $>3\sigma$ certainty. Solid and dotted lines represents the mean abundance and $\pm1\sigma$ abundance values, respectively.  The interquartile range (IQR) of the distribution is shown inset. } 
\label{fig:direct_reg}
\end{figure*}

Chemical abundances were calculated following the methodology outlined in \citet{James:2015,James:2017}. To summarize here, chemical abundances are calculated using the `direct method', where abundance measurements are based on the physical conditions of the gas (i.e.,utilizing electron temperature, \elt, and electron density, \eld) and extinction corrected line fluxes. Each \hii\ region is modelled by three separate ionization zones (low, medium, and high), and the abundance calculations for ions within each zone are made using the temperature within the respective zone. We refer the reader to \citet{James:2017} for further details. The only difference between the methods used here and that of \citet{James:2017} is that due to MUSE's restricted wavelength coverage in the blue, here we use the \fsiii~$\lambda$6312 auroral line to derive the electron temperature (\elt) rather than \foiii$\lambda$4363. Despite its less common usage compared to \foiii ,  the \fsiii\ emission line ratio \fsiii~$\lambda$9071/$\lambda$6312 has been demonstrated to be more accurate in deriving electron temperatures and subsequently \hii\ region abundances \citep{Berg:2015}.   The \eld, \elt, ionic and elemental oxygen abundances, nitrogren abundances, and nitrogen-to-oxygen abundance ratios  for each \hii\ region are shown in Table~\ref{tab:abunds}. For three \hii\ regions (IDs 11, 15, and 26) it was necessary to adopt the \elt\ of the closest neighbouring \hii\ region. Despite \fsiii~$\lambda$6312 being detected at a $>3\sigma$ level in these regions, \elt(\fsiii) values were found to be unfeasibly high ($>$30,000~K)\footnote{Such extremely low oxygen abundance values were not supported by the \fnii /\ha\ line ratio \citep{Pettini:2004} and weak \hei\ lines (which should be relatively strong in comparison to metal lines in the XMP regime).}, which upon investigation may be due to \fsiii~$\lambda$9071 having an asymmetric line profile due to poor sky subtraction. The \fsiii~$\lambda$9071 line profile in all other \hii\ regions were inspected and found to be uncontaminated.

The direct method oxygen abundance calculated for each of the individual \hii\ regions is plotted as a function of distance from the central \hii\ region in Figure~\ref{fig:direct_reg}, where the `arm' denomination refers to those shown in Figure~\ref{fig:HIIreg}b. Here we only show the \hii\ region direct method abundances where \elt\ could be derived confidently (i.e.,where \fsiii~$\lambda$6312 was measured with 3$\sigma$ significance) and, as a result, only direct method measurements for the most central, brightest, \hii\ regions are available. Nevertheless, it can be seen that JKB~18 harbours \hii\ regions with a large range in oxygen abundance of 7.3$\lesssim$12+log(O/H)$\lesssim$7.9, with an average of 12+log(O/H)$=7.56\pm0.20$ ($\sim0.08$~\Zsol). A variation of $\sim0.6$~dex is quite significant, given than most dwarf galaxies are thought to be chemically homogeneous with [O/H] variations of only 0.1-0.2~dex \citep[e.g.,][]{Skillman:1989,Kobulnicky:1996,Lee:2006,Kehrig:2008,Croxall:2009,James:2010,Berg:2012,Lagos:2014,Lagos:2016,Kumari:2019a} and small uncertainty (i.e., $\sigma<0.1$). Several regions have oxygen abundances outside the $\pm1~\sigma$ distribution (dotted lines) of $\sim$0.2~dex, all within $\sim$1~kpc of the central \hii\ region. Only Arm 3 shows any kind of discernible gradient in that the metallicity increases steadily with distance from the central \hii\ region. In an attempt to quantify the size of the spread in oxygen abundance, we calculate an interquartile range (IQR) of 0.11.

 In Figure~\ref{fig:direct_reg} we show the radial distribution of N/H and N/O throughout JKB~18, as derived using the direct method. Similarly with O/H, significant variations in both elemental abundance ratios can be seen, with $\sim$0.5~dex spreads in both N/H and N/O, and averages values of $<$12+log(N/H)$>=6.41\pm0.14$ and $<N/O>=-1.95\pm0.13$. We calculate IQRs of 0.23 and 0.19 for [N/H] and [N/O], respectively. Along `Arm 3', variations can be $\sim$0.3~dex in N/H and N/O. 
 Similar sized variations are also seen, however, in O/H of the same `arm' (Fig.~\ref{fig:direct_reg}), which suggests that the differences between mechanisms responsible for N-production (e.g., AGB stars) and O-production (e.g., core collapse SNe) may not be the sole cause of this variation in nitrogen.

Unfortunately, it is difficult to make any firm conclusions on the chemical homogeneity of JKB~18 with such a small number of direct-method abundance measurements. In the following sections, we explore the metallicity distribution using a strong-line calibration for metallicity, which does not rely on the detection of faint auroral lines and instead provides us with metallicity measurements throughout the entirety of JKB~18. While the absolute value of the strong-line methods are somewhat dubious compared to the direct-method abundances - i.e., typically showing $\sim$0.7~dex offsets from direct-method abundances \citep[e.g.,][]{Kewley:2008,Moustakas:2010,Lopez-Sanchez:2012} - we can use strong-line calibrations to gauge \textit{relative} metallicity variations within the ionised gas. 

\begin{figure*}
\centering
\begin{subfigure}{.5\textwidth}
  \centering
  \includegraphics[scale=0.6]{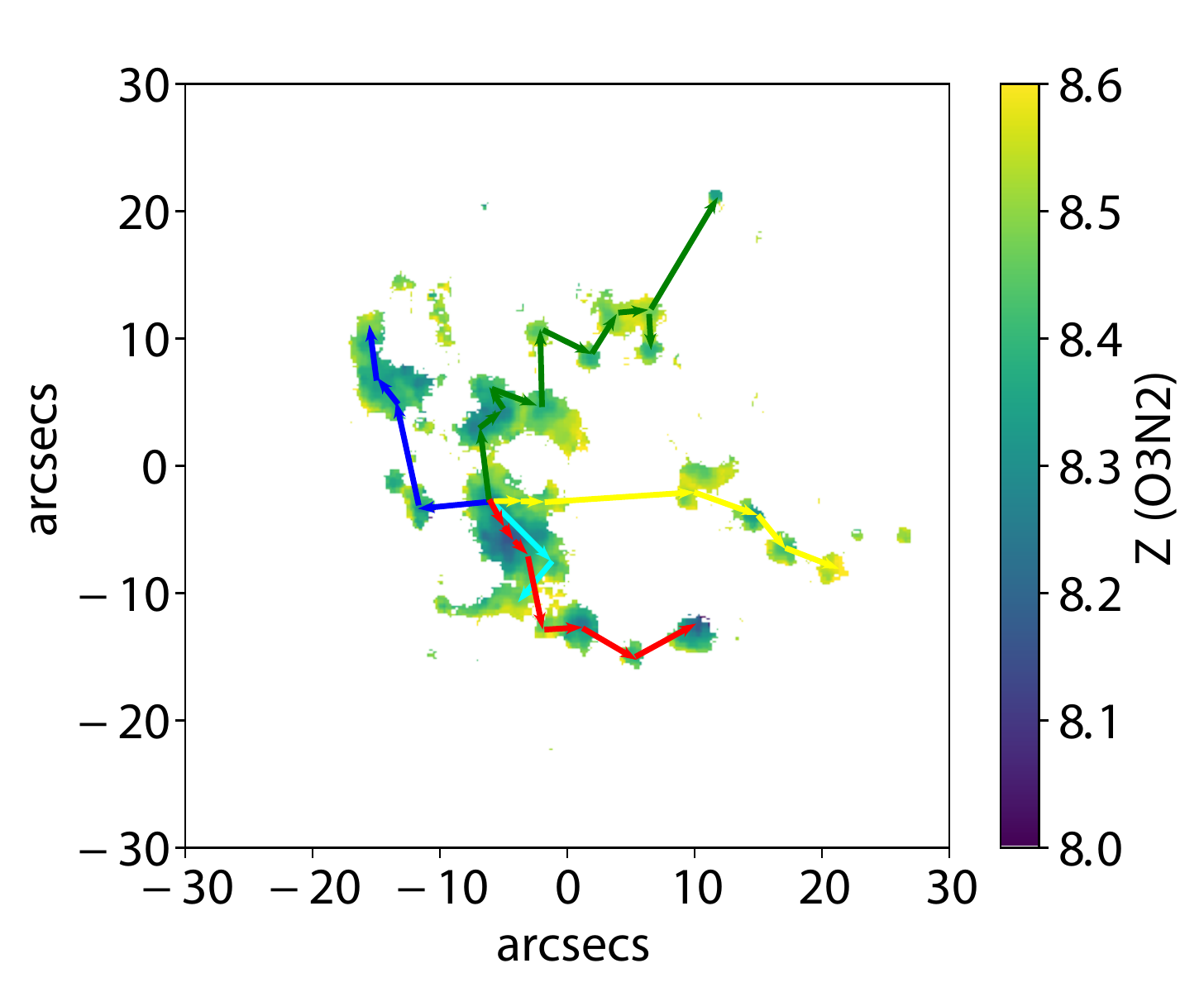}
\end{subfigure}%
\begin{subfigure}{.5\textwidth}
  \centering
  \includegraphics[scale=0.6]{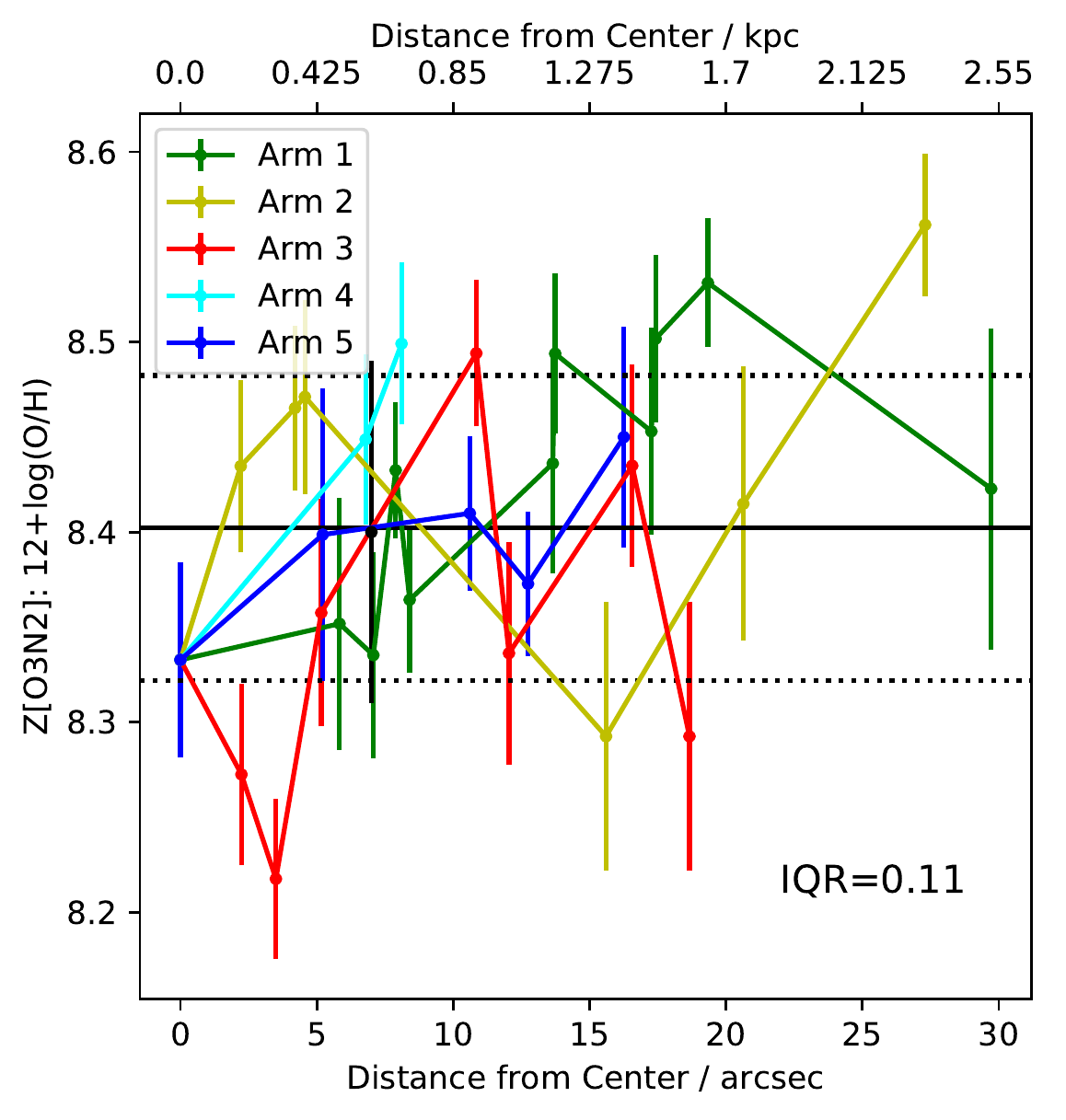}
\end{subfigure}
\caption{\textit{Left-hand panel:} Metallicity map derived from the O3N2 metallicity indicator by \citet{Curti:2017}. Overlaid are the arm substructures identified in Figure~\ref{fig:HIIreg}. \textit{Right-hand panel:} Z[O3N2] metallicity as a function of distance from the central \hii\ region, along the arm substructures shown overlaid in the left-hand panel. Values represent the average Z[O3N2] metallicity measured for all spaxels within that region, along with its standard deviation, as listed in Table~\ref{tab:abunds}. Solid line represents the mean Z[O3N2] metallicity, dotted lines are $\pm1\sigma$. The interquartile range (IQR) of the distribution is shown inset.}
\label{fig:O3N2map}
\end{figure*}

\subsection{Metallicity Map from Strong-Line Diagnostics}\label{sec:met_maps}
 In order to explore the chemical (in)homogeneity throughout JKB~18 on the highest possible spatial resolution afforded by the data, we performed a spatially resolved abundance analysis using the O3N2 metallicity diagnostic (log (\foiii/\hb / \fnii/\ha)) prescribed by \citet{Curti:2017} following the equation: 
\begin{equation}
\rm O3N2 = 0.281-4.765~x_{O3N2}-2.268~x_{O3N2}^2,
\label{eq:o3n2}
\end{equation}
where x$\rm_{O3N2}$ is  oxygen  abundance  normalized  to  the  solar  value in the form 12 + log(O/H)$_{\odot}$ = 8.69 \citep{Asplund:2009}. The Z[O3N2] metallicity map is shown in Figure~\ref{fig:O3N2map}, where the metallicity can be seen to range between 8.1$\lesssim$12+log(O/H)$\lesssim$8.6, an average value of $Z[O3N2]=8.44\pm0.08$ and an IQR of 0.11. As mentioned above, the large $\sim$0.7--0.8~dex systematic discrepancy between the oxygen abundances derived from the direct-method and the strong-line diagnostic are to be expected (with the latter typically showing larger abundances than the former). As such, here we focus only on the relative values of Z[O3N2] between the \hii\ regions, rather than the absolute values.

Small gradients in Z[O3N2] within the gas can be seen, although with no discernable global pattern. To explore these patterns in a more quantitative context, in Figure~\ref{fig:O3N2map} (right) we show the mean Z[O3N2] value (as listed in Table~\ref{tab:abunds}) for each of the \hii\ regions plotted as a function of distance from the central (brightest) \hii\ region following the `arm' structures overlaid in Figure~\ref{fig:O3N2map} (left). Echoing the Z[O3N2] map, the \hii\ region Z[O3N2] metallicities behave in almost a random nature - rather than there being gradients along any of the arms, the metallicity instead fluctuates without any sense of order. Over-plotted on Figure~\ref{fig:O3N2map} (right) are the mean and standard deviation of all the Z[O3N2] values measured within the regions (solid and dotted lines, respectively). There are several \hii\ regions that have metallicity values outside the $1\sigma$ distribution, although the deviations themselves are relatively small ($\sim0.1$~dex). It should be noted that if we were to also include the systematic uncertainties on the Z[O3N2] metallicity calibrator itself (0.09~dex) then the majority of \hii\ regions would have Z[O3N2] values within the 1~$\sigma$ distribution. We present an in-depth discussion pertaining to the possible chemical homogeneity of JKB~18, and dwarf galaxies in general, in given the following sections.

Finally, it is important to note that since the Z[O3N2] metallicity diagnostic is known to have a dependency on both the ionization parameter ($U$) and the nitrogen-to-oxygen ratio of the gas \citep[see e.g.,][ and references therein]{Maiolino:2019}, it would have been beneficial to explore whether the inhomogeneities seen in the Z[O3N2] distribution also exist in strong-line metallicity diagnostics that do not have such dependencies. The only two diagnostics that satisfy this requirement would be the `R23' and `O3S2' diagnostics, which have only a secondary dependence on $U$ and no dependence on nitrogen. Unfortunately, neither of these diagnostics can be applied here. We are unable to compute Z[R23] because the required \foii~$\lambda\lambda$3727, 3729 doublet lies beyond MUSE's wavelength range and the low-metallicity `branch' of the O3S2 distribution is only valid for 12+log(O/H)$>$7.6 \citep{Curti:2017,Kumari:2019b}, which excludes the vast majority of \hii\ regions within JKB~18.

\begin{center}
\begin{table*}
\caption{\ha\ luminosity and star-formation rates for each individual \hii\ region identified throughout JKB~18 }\label{tab:sfr}
\begin{center}
\begin{footnotesize}
\begin{tabular}{|c|cc|}
\hline
\hii\ Region ID & $L$(\ha) ($\times10^{36}$ erg\,s$^{-1}$) & SFR(\ha) ($\times10^{-2}$ \Msol \,yr$^{-1}$) \\
\hline
0 & 129.46 $\pm$   0.16 &   0.47 $\pm$   0.11 \\
1 &  91.51 $\pm$   0.18 &   0.38 $\pm$   0.09 \\
2 &  72.99 $\pm$   0.12 &   0.33 $\pm$   0.08 \\
3 &  82.11 $\pm$   0.14 &   0.36 $\pm$   0.08 \\
4 &  55.30 $\pm$   0.10 &   0.28 $\pm$   0.07 \\
5 &  78.23 $\pm$   0.13 &   0.35 $\pm$   0.08 \\
6 &  43.97 $\pm$   0.15 &   0.24 $\pm$   0.06 \\
7 &  57.96 $\pm$   0.13 &   0.29 $\pm$   0.07 \\
8 &  23.00 $\pm$   0.09 &   0.16 $\pm$   0.04 \\
9 &  28.83 $\pm$   0.09 &   0.19 $\pm$   0.05 \\
10 &  35.13 $\pm$   0.10 &   0.21 $\pm$   0.05 \\
11 &  24.66 $\pm$   0.10 &   0.17 $\pm$   0.04 \\
12 &  13.81 $\pm$   0.10 &   0.12 $\pm$   0.04 \\
13 &  13.26 $\pm$   0.06 &   0.12 $\pm$   0.04 \\
14 &  30.42 $\pm$   0.13 &   0.19 $\pm$   0.05 \\
15 &  23.29 $\pm$   0.10 &   0.16 $\pm$   0.04 \\
16 &  22.54 $\pm$   0.07 &   0.16 $\pm$   0.04 \\
17 &  17.24 $\pm$   0.09 &   0.14 $\pm$   0.04 \\
18 &  14.25 $\pm$   0.13 &   0.12 $\pm$   0.04 \\
19 &  24.39 $\pm$   0.10 &   0.17 $\pm$   0.04 \\
20 &  22.48 $\pm$   0.09 &   0.16 $\pm$   0.04 \\
21 &  19.82 $\pm$   0.18 &   0.15 $\pm$   0.04 \\
22 &   9.10 $\pm$   0.05 &   0.09 $\pm$   0.03 \\
23 &  14.38 $\pm$   0.08 &   0.12 $\pm$   0.04 \\
24 &  19.71 $\pm$   0.18 &   0.15 $\pm$   0.04 \\
25 &  14.25 $\pm$   0.14 &   0.12 $\pm$   0.04 \\
26 &  19.03 $\pm$   0.11 &   0.14 $\pm$   0.04 \\
27 &  20.40 $\pm$   0.14 &   0.15 $\pm$   0.04 \\
28 &   8.58 $\pm$   0.08 &   0.09 $\pm$   0.03 \\
29 &  21.48 $\pm$   0.14 &   0.16 $\pm$   0.04 \\
\hline
\end{tabular}
\label{tab:SFR}
\end{footnotesize}
\end{center}

\end{table*}
\end{center}

\section{Discussion}\label{sec:disc}
In the following section we attempt to put our findings into context by discussing the spatial variation of metals within three frames of reference: dwarf galaxies in general, JKB~18 alone, and galaxy-scale simulations.

\subsection{How chemically homogeneous \textit{are} dwarf galaxies?}
Considering the fact that spatially resolved chemical abundance studies have only become prevalent since the advent of IFUs on almost all of the large ground-based observatories (e.g., VLT, Keck, Gemini), and large-scale IFUs only being brought online during the past $\sim$5 years,  we do not have an overwhelming body of literature to draw from on this subject. While chemical abundance mapping of large-scale spiral galaxies is very well populated in the literature, those concentrating on dwarf galaxies is not. From the studies that have looked into this so-far, the majority of studies suggests that ionized gas on scales of $>$100~pc is chemically homogeneous \citep[e.g.,][]{Kehrig:2008,James:2010,Perez-Montero:2011,Kehrig:2013,Lagos:2014,Lagos:2016,Kumari:2019a} with spreads of $<0.2$~dex within the uncertainties. However, this is not always the case: 
\begin{itemize}
\item In Haro~11, a nearby (83.6~Mpc) BCD galaxy $\sim10^9$~\Msol\ in size, chemical abundance maps from VLT-FLAMES revealed variations in O/H of $\sim$0.5~dex in size between its three star-forming knots separated by $\sim$1~kpc \citep{James:2013a}. In this particular galaxy, depleted O/H and kinematical  signatures of outflowing gas in Knot~C suggested an outflow of metal-enriched gas due to SNe. 
\item In another BCD, UM~448 ($D\sim$76~Mpc) differences of $\sim$0.4~dex in O/H were found between the broad and narrow component emission in the main body of the galaxy, along with a 0.6~kpc$^2$ region of enhanced N/O \citep{James:2013b}. Here, the disturbed morphology and dynamics of the gas in this particular region are reminiscent of interaction-induced inflow of metal-poor gas.  A similar scenario can be applied to NGC~4449, another nearby BCD galaxy, where a localized area of depleted oxygen abundance (by 0.5~dex) aligns spatially with a region of increased star-formation in the central $\sim0.5$~kpc of the galaxy, most likely due to metal-poor gas accretion due to an ongoing merger event \citep{Kumari:2017}.
\item In BCD UM~462, the three main regions of star-formation show decreasing metallicities (by $\sim$0.1~dex between regions) alongside increasing stellar population ages and star-formation rates (SFR) that decreased by a factor of ten \citep{James:2010}. While the variations in chemical abundance are small here, such patterns imply triggered or propagated star-formation, where star-forming regions retain signatures of their chemically processed gas due to short mixing timescales. However, it should also be noted that UM~462 may have been tidally disrupted due to an interaction with its neighbour, UM 461, thereby promoting efficient flattening of its metallicity and dilution by low-metallicity gas infalling into the galaxy centre \citep{Lagos:2018}.
\item NGC~5253 and Mrk~996 are two star-forming dwarf galaxies famous for showing high levels of nitrogen over-abundance, despite uniform oxygen abundance distributions. In the case of NGC~5253 ($D\sim$3.8~Mpc), a 0.5~dex enhancement in N/H is seen in the central $\sim$50~pc region, coincident with a supernebula containing nitrogen rich Wolf Rayet (WR) stars \citep[][and references therein]{Westmoquette:2013}. A similar story is found for Mrk~996, where a N/O enhancement of $\sim$1.3~dex is seen in the broad component emission emanating from the central $180\times250$~pc region \citep{James:2009}, coincident with a population of evolved massive stars such as WNL-type or luminous blue variables. Another recent addition to the `N-enhanced dwarf galaxies group'  is NGC~4670 \citep{Kumari:2018}, where a $\sim$0.2~dex enhancement in N/O aligns spatially with a population of WR stars. 
\item Enhancements in N/H are not always due to enrichment from WR-stars. In the case of Tol~2146-391, a $\sim$0.5~dex variation in N/H ($\sim$0.2~dex outside the 1$\sigma$ distribution) is seen across a distance of $\sim1$~kpc from the nucleus, despite a lack of WR signatures and other elements appearing well mixed \citep{Lagos:2012}. The authors attribute this radial pattern to heavy elements that were produced during the previous burst of star-formation and which are currently dispersed by the expansion in the ISM of starburst-driven supershells.
\item As discussed in \citet{Bresolin:2019}, there are a handful of dwarf-irregular galaxies that actually have \textit{well-ordered} gradients that are comparable to spiral galaxies. For example, NGC~6822 \citep{Lee:2006}, NGC~1705 \citep{Annibali:2015}, NGC~4449 \citep{Annibali:2017}, and DDO~68 \citep{Annibali:2019} each show chemical abundance gradients, despite the absence of spiral structures which were deemed necessary for gradients below $M_B\simeq17$ by \citet{Edmunds:1993}. \citet{Bresolin:2019} link the existence of these gradients to recent enhancements in their star-formation activity, where metal mixing timescales are longer than the time between bursts of star-formation.
\end{itemize}

From each of these cases, it is apparent how IFU studies, and spatially-resolved studies in general, have the advantage of not only enabling chemical abundance mapping itself, but also putting such maps into an evolutionary context via a multitude of additional spatially derived properties (e.g., kinematics, ionising population age, and morphology). As such, we can attribute the chemical inhomogeneities to either (i) outflows of metal-enriched gas from SNe, (ii) the accretion of metal-poor gas due to interactions/mergers, (iii) self-enrichment from winds of massive stars, or (iv) observing bursts of star-formation at times within the mixing timescale. While the results described above only offer a handful of chemically inhomogeneous dwarf galaxies, they do demonstrate that such cases \textit{do} exist and assuming chemically homogeneous gas to exist across large spatial scales throughout all low-metallicity star-forming galaxies (e.g., such as those we would expect to observe in the early universe) would be detrimental to studies of galaxy formation and evolution.

Finally, it should be noted that the spatial scale of chemical mapping studies can also play a large role in determining the chemical inhomogeneity of dwarf galaxies, both with regards to the spatial resolution of the spectra and the size of the field over which measurements are made. While increasing the spatial resolution can both dilute metallicity gradients and remove any sharp structural features \citep[e.g.,][]{Kobulnicky:1998,Yuan:2013,Mast:2014, James:2016}, high spatial resolution chemical homogeneity studies over limited portions of a galaxy are also restricted in their conclusion \citep[e.g.,][]{Kumari:2017} given that large scale variations are not expected within the scales of \hii\ regions ($<100-200$~pc). Studies exploring chemical homogeneity on extremely small spatial scales of $<10$~pc are very rare, mostly because it became possible to achieve such unprecedented spatial scales only since the advent of VLT/MUSE. One such study probing metallicity on sub-pc scales was performed by \citet{James:2015} using HST/WFC3 narrow band images of Mrk~71 (a nearby BCD galaxy at $D\sim$3.4~Mpc) to derive metallicity maps on scales of only $\sim$2~pc. While this galaxy did reveal chemical variations of 0.2--0.3~dex on scales $<$50~pc, the study found that applying strong-line metallicity diagnostics on spatial scales smaller than an \hii\ region can be problematic in that they may effectively `break down' as their design is based on light integrated over entire or multiple \hii\ regions.

\subsection{Is JKB~18 Chemically Inhomogeneous?}\label{sec:inhomo_disc}
Overall, the question of whether the gas throughout JKB~18 is chemically homogeneous or not, is somewhat difficult to quantify. Perhaps first the question of \textit{how} we quantify chemical (in)homogeneity becomes significant. Several previous studies have used the 1$\sigma$ distribution as a boundary beyond which any chemical variations are considered 'inhomogeneous' \citep[e.g.,][]{James:2016,Kumari:2017,Kumari:2019a}. However, the \textit{size} of the 1$\sigma$ distribution itself should also be taken into account, such that a large spread in oxygen abundances also deems gas to be chemically inhomogeneous. If this were not the case, a significant fraction of shallow metallicity gradients would not be considered gradients, given the uncertainties on the measurements. Also, using only deviations outside standard deviation of the distribution to define inhomogeneity assumes that distribution is normal, or Gauassian, which is not always the case. In fact, this was indeed the test assumed in a chemical homogeneity study by \citet{Perez-Montero:2011}, where only distributions that could be fitted with a Gaussian function were deemed homogeneous. As such, an additional useful measure of (in)homogeneity could also be the interquartile range of the distribution in that outlier values are not taken into consideration. In the case of JKB~18, on the one hand a large range in oxygen abundances of $\sim$0.5--0.6~dex is observed between individual \hii\ regions, using both the direct method and the $O3N2$ strong-line metallicity indicator. However, when we consider these variations in context of the mean and standard deviation of the oxygen abundance distribution, variations of only $\sim$0.1~dex are seen outside the 1$\sigma$ distribution - but the 1$\sigma$ distribution in itself is $\sim$0.2~dex in size. The IQR is somewhat smaller, of the order $\sim$0.1~dex, suggesting that a large number of outlier values are present. If we consider much smaller spatial scales, i.e., looking at spaxel-to-spaxel variations in $Z[O3N2]$ (Figure~\ref{fig:O3N2map}) then variations of $\sim0.2-0.3$~dex are seen within regions the size of typical \hii\ regions. The 1$\sigma$ distribution on this spatial scale is $\sim$0.1~dex whereas the IQR=0.13. 
\begin{center}
  \centering
  \includegraphics[scale=0.6]{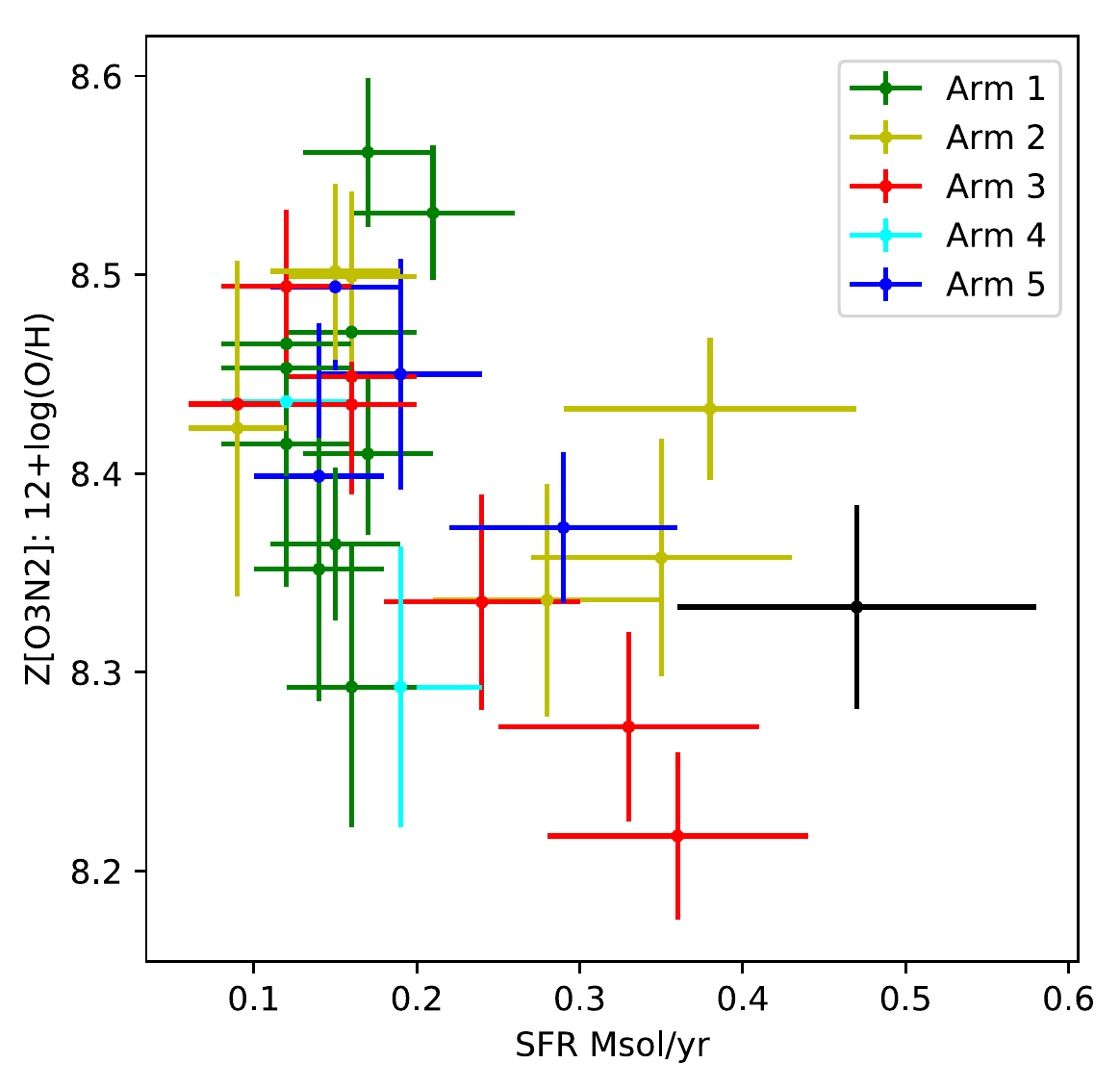}
  \captionof{figure}{Z[O3N2] as a function of SFR (as listed in Table~\ref{tab:sfr}). No correlation can be seen between the two properties, suggesting that inflow of metal-poor gas is not the dominant mechanism influencing the metallicity of the individual \hii\ region. The central \hii\ region is shown in black.}
  \label{fig:Z_SFR}
\end{center}

\begin{center}
\includegraphics[scale=0.6]{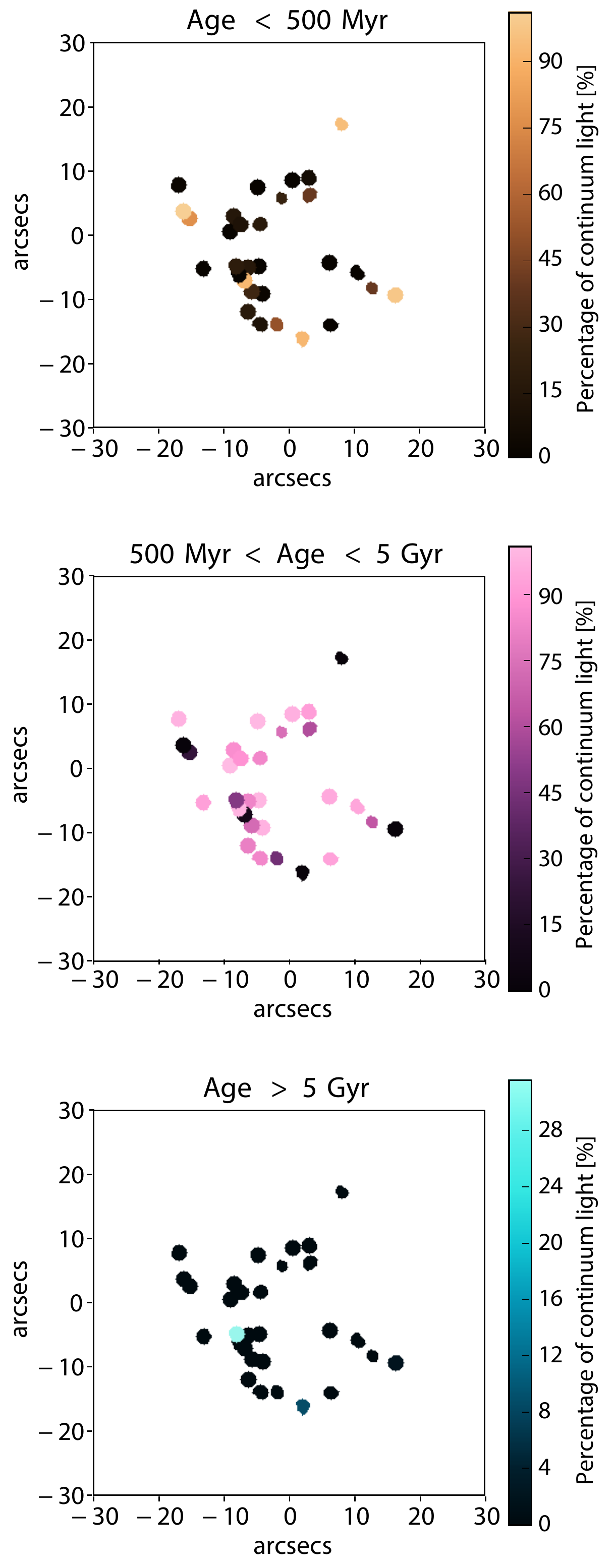}
\captionof{figure}{Age of the stellar population in each \hii\ region, as identified in Section~\ref{sec:HIIregs}. The three panels show the percentage of light contributing to the stellar continuum from three separate age bins: $<500$~Myr, $500~Myr < Age <5$~Gyr, and $>5$~Gyr.} 
\label{fig:stellar_age}
\end{center} 

Overall, given the significant variation in oxygen abundance, both with regards to the size of the 1$\sigma$ distribution and the fact that several \hii\ regions lie outside this distribution, we find JKB~18 to be chemically inhomogeneous.

While only small temperature variations seen \textit{within} JKB~18's individual \hii\ regions are to be expected, what would be the cause of 0.5--0.6~dex variations in oxygen abundance \textit{between} \hii\ regions?  To explore the cause of the \hii\ region metallicity distribution of Figure~\ref{fig:O3N2map} further, in Figure~\ref{fig:Z_SFR} we show the Z[O3N2] metallicities as a function of SFR. SFR values for each individual \hii\ region are listed in Table~\ref{tab:sfr} and were calculated following the method outlined in \citet{James:2017} in that we follow the prescription of \citet{Lee:2009} because all individual \hii\ regions are below the low-luminosity threshold of $L($\ha$)<2.5\times10^{39}$ erg\,s$^{-1}$.\footnote{This prescription is essentially a recalibration the \citet{Kennicutt:1998} relation to account for the underprediction of SFRs by \ha\ compared to those from FUV fluxes within the low-luminosity regime, under the assumption that the FUV traces the SFR in dwarf galaxies more robustly.} We find an average star-formation rate and 1$\sigma$ distribution of $0.19\pm0.09\times10^{-2}$~\Msol/yr across the 30 \hii\ regions identified in Figure~\ref{fig:HIIreg}. If there were inflows of metal poor gas affecting the metallicity of an \hii\ region, then we would expect to see an anti-correlation between SFR and Z[O3N2] - i.e., as metal-poor gas flows into a region, both the metal content of the gas becomes diluted and star-formation is triggered. This effect is commonly seen in XMP and `tadpole' galaxies, as shown in \citet{Sanchez-Almeida:2014b}. In order to understand whether such a correlation exists here, a Pearson correlation test was performed on the parameters shown in Figure~\ref{fig:Z_SFR}, which resulted in a Pearson correlation coefficient of $\rho=-0.43$. As such, only a weak negative correlation exists between Z and SFR, suggesting that in inflow of metal-poor gas cannot entirely be responsible for the decreased oxygen abundance in seen in some regions of JKB~18.

If instead, mixing timescales were sufficiently short enough that the metal content of the gas surrounding the stars reflected the evolutionary age of the material ejected by the star \citep[i.e., as proposed by][]{Kobulnicky:1997}, then we would expect the metallicity of the \hii\ region to increase with stellar age. In Figure~\ref{fig:stellar_age} we show the contribution of stellar populations within each \hii\ region, separated in three age bins ($<500$~Myr, $500$~Myr$ < Age <5$~Gyr, and $>5$~Gyr). The percentage contributions were calculated from the best-fit parameters to the stellar continuum of the integrated spectrum of each \hii\ region using STARLIGHT \citep{CidFernandes:2005} implemented within IFUANAL (Section~\ref{sec:met_maps}).  In short, the spectral-fitting code STARLIGHT allows us to fit the continuum of a spectrum (where the emission lines are masked) with a model composed of spectral components from a pre-defined set of 45 base spectra from \citep{Bruzual:2003} corresponding to 3 metallicities and 15 different ages, and outputs the light-fraction (contribution in \%) of each base spectrum among various parameters. The contribution of stellar populations shown in Figure \ref{fig:stellar_age} corresponds to the sum of contribution of the base spectra corresponding to a given stellar age bin. We find that the majority of \hii\ regions have their largest percentage contribution within the $500$~Myr$ < Age <5$~Gyr age range. There are several \hii\ regions whose continuum mostly indicates a young, $<500~Myr$ population, which appear to lie mostly in the outskirts of the galaxy. A few \hii\ regions also contain stellar populations older than 5 Gyr (bottom panel) though their fraction is $<5\%$ (with the exception of the central \hii\ region), signifying that these \hii\ regions have gone through episodes of star-formation in the past which have resulted in an underlying older population.     

Despite the lack of metallicity gradients in JKB~18, the distribution of stellar population age suggests that JKB~18 may have undergone inside-out growth, such that older stellar populations lie in the center and younger populations in the outskirts of the galaxy. This can perhaps be seen more clearly in Figure~\ref{fig:age_dist}, where we plot the light-weighted average age of each \hii\ region\footnote{Estimated via the light-fraction of base spectra output from STARLIGHT as described above.} of the stellar population as a function of radius from the central \hii\ region - the average age of the populations within the central $\sim$0.5~kpc of the galaxy are mostly $\gtrsim$2~Gyr in age, which then plateaus into a large age distribution $\sim$0.5--1.8~kpc from the center of 0.1--2~Gyr, with the outermost \hii\ regions ($>1.8$~kpc) all with ages less than $0.5$~Gyr. In Figure~\ref{fig:Z_age} we plot the light-weighted average stellar population ages as a function of Z[O3N2]. Once again, we see no correlation between the two properties, suggesting that long mixing timescales and thus the evolutionary stage of the star are not affecting the metal content of the surrounding gas. 

Finally, the increased levels of N/H and N/O seen in Figure~\ref{fig:direct_reg} may be due to self-enrichment from WR stars within those particular regions (e.g., those along Arm 3). However, upon inspection of the individual spectra, WR features at $\sim$4650~\AA\ and $\sim$5800~\AA\ were not present. 

\begin{center}
\includegraphics[scale=0.6]{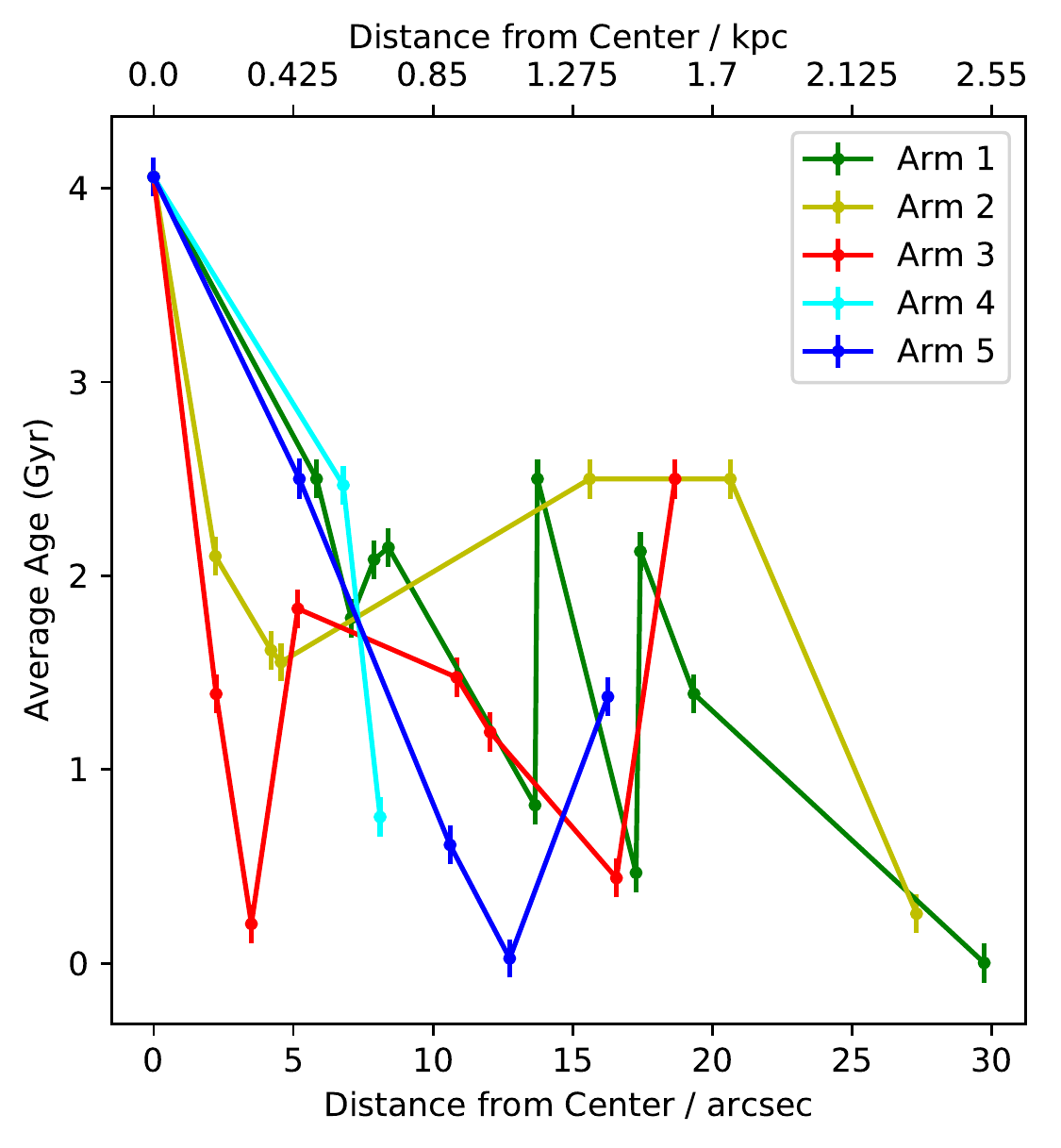}
\captionof{figure}{Light-weighted average age of the stellar population (discussed in Section~\ref{sec:inhomo_disc}) as a function of distance from the central \hii\ region, along the arm substructures shown in Figure~\ref{fig:HIIreg}.} 
\label{fig:age_dist}
\end{center}

\begin{center}
\includegraphics[scale=0.6]{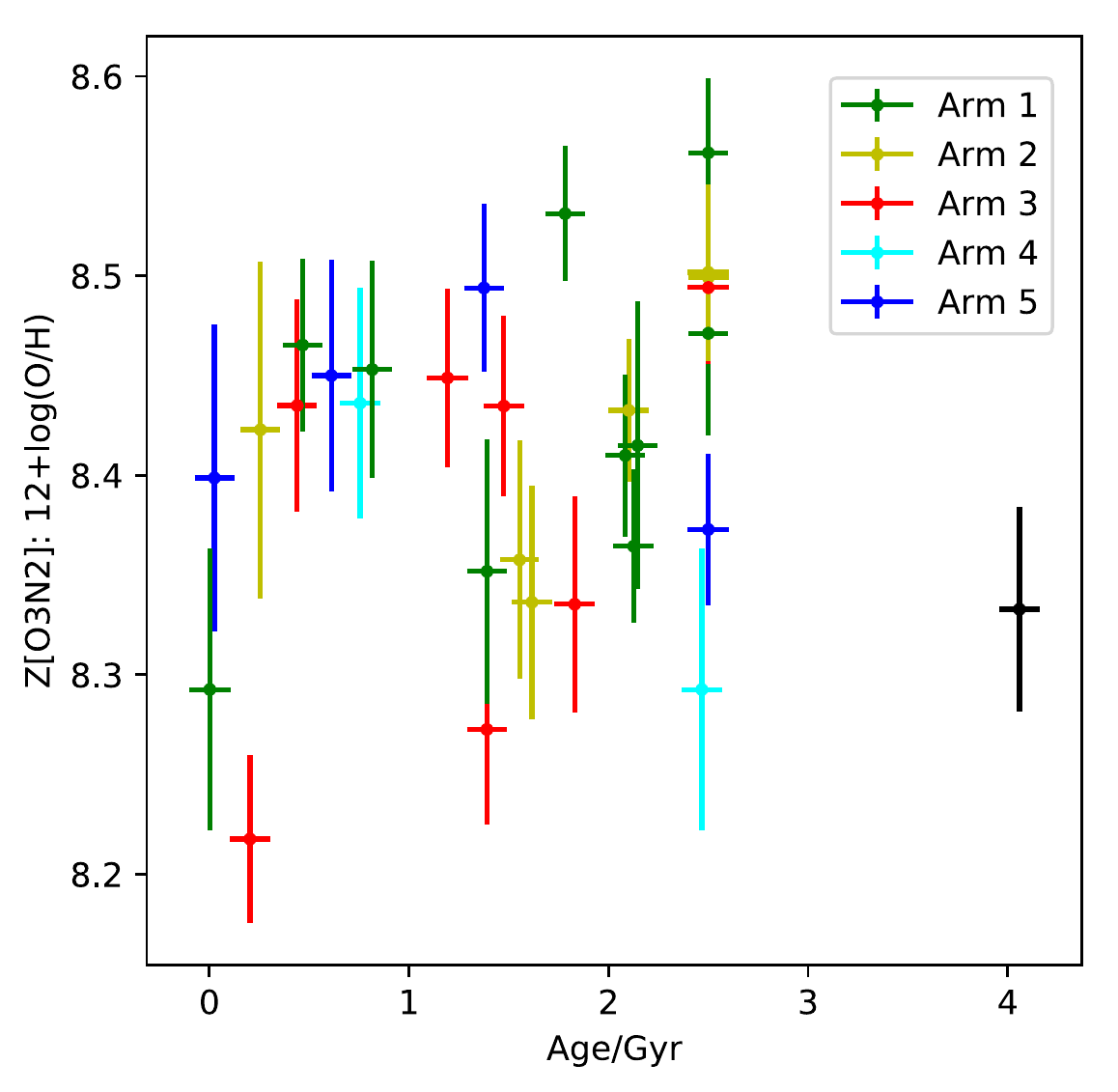}
\captionof{figure}{Z[O3N2] as a function of the light-weighted average age of the stellar population, for each \hii\ region in JKB~18 (derived from Fig.~\ref{fig:stellar_age}). The central \hii\ region is shown in black. The uncertainties on stellar age correspond to the standard error on the light-weighted age, estimated from the light-fractions output from the STARLIGHT, while the uncertainties on metallicity correspond to the standard deviation of metallicities of all pixels within an \hii\ region. No correlation is seen between the age of \hii\ regions and its metallicity, suggesting short mixing timescales and/or efficient ejection of metals from the dwarf galaxy.} 
\label{fig:Z_age}
\end{center} 

The fluctuations in metallicity seen in Figures~\ref{fig:direct_reg} and \ref{fig:O3N2map}, and the lack of correlation between metallicity and SFR or stellar population age suggest two properties about this system: (i) enrichment events (e.g., SNe or stellar ejecta) are currently ongoing and occurring at different time intervals, resulting in a dispersed metal content and (ii) the metals ejected by the stars within the individual \hii\ regions have been removed from the  galaxy and are yet to cool into the warm ionized gas phase (i.e., which is visible with the MUSE data presented here). We explore these two items in more detail in the context of dwarf galaxy evolution in the following section.

\begin{figure*}
\includegraphics[scale=0.4]{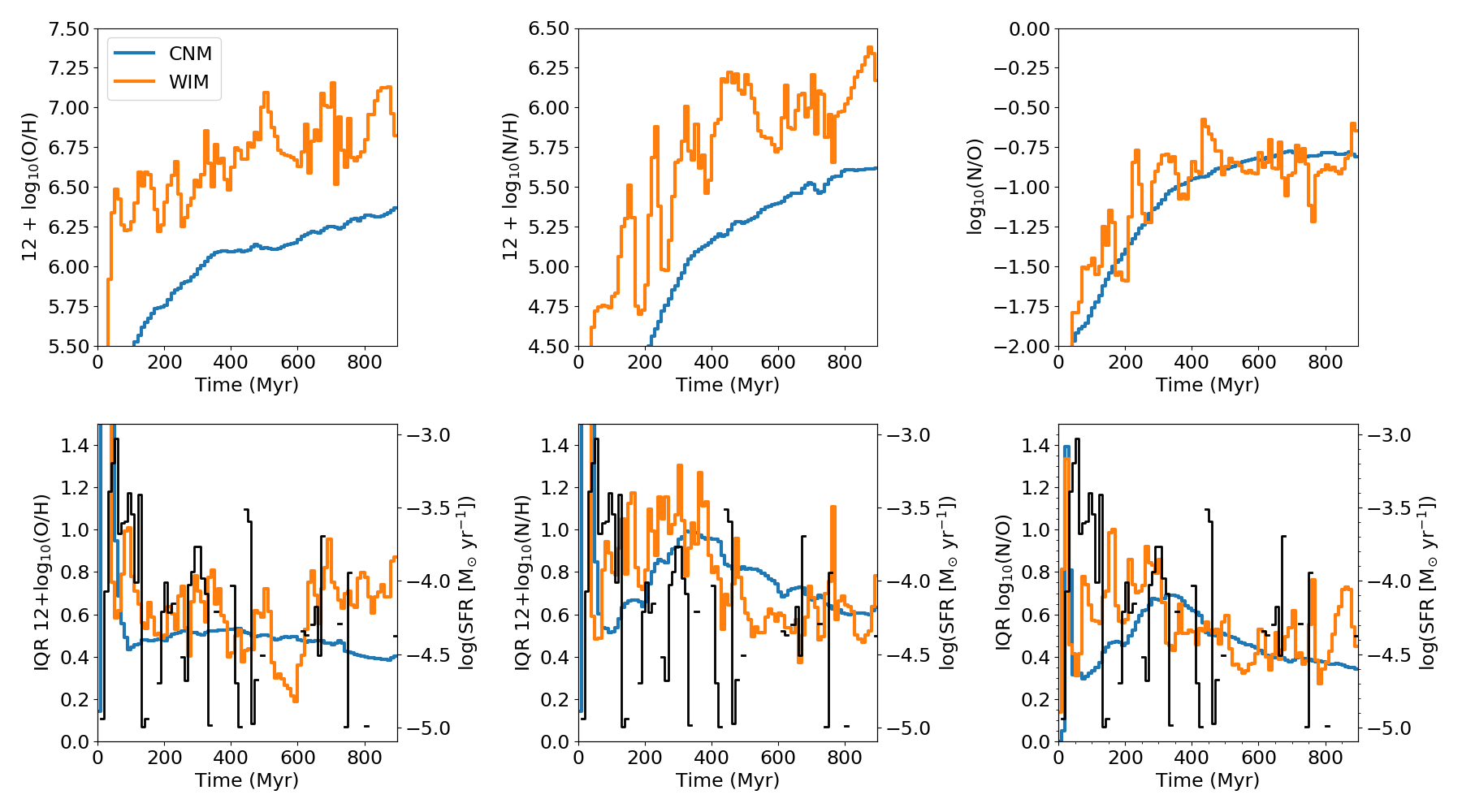}
\caption{A quantification of chemical homogeneity throughout the Leo~P-like simulated dwarf galaxy presented in \citet{Emerick:2019}. The top panels show the median [O/H], [N/H] and [N/O] levels for the cold neutral medium ($T<100$~K) and warm ionized medium ($10^4< T(K) < 10^{5.5}$) as a function of time throughout the simulation. The bottom panels show the inter-quartile range (IQR) for the same distributions as a function of time, with the SFR overlaid in black.  All curves are averaged over 10~Myr intervals; gaps in the SFR are physical, and are periods with no star formation.} 
\label{fig:sim_spreads}
\end{figure*} 

\subsection{What can simulations tell us about dwarf galaxy chemical homogeneity?}\label{sec:sim_disc}

Recent works have begin to include sufficient, detailed physics to capture metal mixing in dwarf galaxies \citep[e.g][]{Shen2010,Shen2013,Few:2012,Brook2014,Revaz2016,Su2017a,Jeon2017,Hirai2017,Corlies:2018,Cote:2018,Escala2018}. But these works generally focus on stellar metallicity distributions as opposed to the type gas-phase distributions studied for JKB18. We turn to one of these recent models for insights into the origin of the observed gas-phase abundance distributions in JKB18. The simulations of \citet{Emerick:2019} follow the star formation, stellar feedback, and multi-phase chemical enrichment of 15 metal species of a Leo~P like dwarf galaxy \citep[$M_{HI}=8\times10^5$~\Msol, $M_{\star}=6\times10^5$, 12+log(O/H)=7.17][]{Giovanelli:2013,Skillman:2013,McQuinn:2015a} to high resolution (1.8~pc) for individual stars. While lower mass than JKB18, the properties of this simulated galaxy make it somewhat relatable to that of blue diffuse dwarf galaxies \cite[whose discovery parameters were based on the morphological properties of Leo~P,][]{James:2016}. We utilize the outcomes of this model (referred to hereafter as the `Leo~P model') to help interpret our findings on JKB~18 and other chemically inhomogeneous dwarf galaxies (e.g., NGC~5253, Mrk~996, NGC~4670). However, it should be noted that since this simulation represents only a single specific system, with a particularly low mass, the results of the simulation should be interpreted as showing \textit{possible} rather than inevitable scenarios for the  mixing/retention/removal of metals in dwarf galaxies. 

First, we note that the Leo~P model demonstrates that mixing experienced by individual metal elements is dependent on their nucleosynthetic origin \citep{Emerick:2018b}. For example, elements dominated by AGB wind enrichment (e.g., N and Ba at the low metallicities in the model) mix less efficiently than elements dominated by supernova enrichment (e.g., $\alpha$ elements and Fe). This would imply larger inhomogeneities in N/H compared to O/H, which may give rise to regions / pockets of elevated N/O  in dwarf galaxies with AGB-driven N enrichment. As discussed above, this is precisely what we have seen in several dwarf galaxies (e.g., NGC~5253, Mrk~996, and NGC~4670).  

Examining the spatial distribution of metals in the Leo~P simulation, we explore the chemical homogeneity as a function of time for both the CNM ($T\lesssim100$~K) and WIM ($10^4\lesssim T(K) \lesssim 10^{5.5}$) in Figure~\ref{fig:sim_spreads}.\footnote{We note that this model was initialized with zero initial O and N abundance, no initial background stellar population, and adopts metallicity dependent stellar yields assuming a total metallicity comparable to that of Leo~P. For the adopted yield tables \citep{Ritter:2018}, significant N is produced in the AGB phases of 5-7~M$_{\odot}$ stars; no model for N production in Wolf-Rayet stars is included.} The top panels shows the median abundances in the CNM and WIM over time, while the bottom panels show the interquartile range (IQR) of the gas-phase abundance distributions. Examining the IQRs, we can see that the cold, neutral gas (blue) is more homogeneous than the warm, ionized gas (orange). While the median O and N abundances of the WIM fluctuates by $\sim$0.5~dex on tens of Myr timescales (top panels), the CNM evolution is gradual and (nearly) monotonic. The WIM fluctuations are driven by cycles of star formation and stellar feedback, which both generates and enriches the WIM with new metals. To see this, we overlay the SFR (black) in the bottom panels. These fluctuations are also seen clearly in the IQR of the WIM, which can range between a minimum of $\sim$0.2~dex up to $\sim$0.9~dex in O/H and $\sim$1.2~dex for N/H. The IQR of the CNM is comparitavely steady for O, but grows (then decreases) for a period for N as N production from AGB winds "turns on" after a few hundred Myr. These trends are reflected in the N/O abundance ratio. This suggests that one possible explanation for large scale variations in O/H and N/H is a recent increase in SFR. As discussed above, a similar hypothesis is also used by \citet{Bresolin:2019} to explain abundance gradients in dwarf irregular galaxies. It is possible that JKB~18 too is undergoing a period of increased star formation activity, resulting in a broadened distribution of metal content between \hii\ regions. This could be tested with similar observations of galaxies with a range of measured star formation rates.

Overall, the Leo~P simulation demonstrates that, like JKB~18, star-forming dwarf galaxies can indeed be chemically \textit{in}homogeneous (in their WIM), and the exact size of that inhomogeneity may depend highly on the timing of our observations with respect to star-formation activity. However, by comparison of their metallicity distributions, the $\sim$0.1~dex IQR in O/H and $\sim$0.2~dex IQRs for N/H and N/O of JKB~18 (Figures~~\ref{fig:direct_reg} and \ref{fig:O3N2map}) suggests that JKB~18 is more homogeneous than the WIM gas simulated in the Leo~P model. This is perhaps not surprising - the amplitude of the change in chemical homogeneity should be smaller for a galaxy like JKB~18 for several reasons. Firstly, JKB~18 is 2 orders of magnitude larger than Leo~P in stellar mass, corresponding to a deeper dark matter potential, and therefore is more likely to retain a larger percentage of the metals from each SNe event compared to Leo~P. Secondly, JKB~18 is a factor of $\sim$2.5 more metal rich than Leo~P and has a higher current level of star-formation by an order of magnitude \citep[based on Leo~P's \ha\ SFR of $4.3\times 10^{-5}$~\Msol/yr,][]{McQuinn:2015a}. 
As such, there is a greater likelihood that (i) much of the galaxy has been directly enriched by each type of event (e.g., core-collapse or Type Ia SNe, stellar ejecta from AGB stars etc) and (ii) the metal mass from such an event is small compared to the total metal mass of the galaxy. This first effect significantly reduces the magnitude of galaxy-wide abundance spreads \citep{Emerick:2019b}, and the second would reduce the amplitude of the fluctuations that are seen in the WIM of the Leo~P simulation. Finally, the star-formation history in general may have been very different - i.e., rather than the bursty SF experienced by the Leo~P model, JKB~18 may have had a more steady level of recent star-formation.

With regards to the mixing timescales of gas in dwarf galaxies, since we do not find regions of enrichment surrounding JKB~18's \hii\ regions, we can infer that either the nucleosynthetic products from massive stars are either subject to long mixing times scales, or the products are no longer observable (i.e., lying in a different phase) and may remain as such for much longer timescales - perhaps even longer than the lifetime of the \hii\ region, as proposed by \citet{Tenorio-Tagle:1996}. It is most likely that both effects are at play here, given the fact that we expect low mass dwarf galaxies of this kind to only retain 5--20\%\ of their metals and, as demonstrated in the Leo~P simulation, the majority of metals may instead be hosted in the cold neutral gas \citep{Emerick:2018b}. Since mixing timescales are on the order of gigayears in the cold gas phase, one would expect to see large variations in metal abundances, especially for metal species resulting from sources less energetic than SNe (e.g., AGB stellar winds). This is indeed the case for the Leo~P simulation, where the CNM does not show fluctuation in its O/H variation (Figure~\ref{fig:sim_spreads}) and instead remains constant at $\sim$0.4~dex, while N/H does change by $\sim$0.5~dex over 900~Myr. However, this may not be the case for JKB~18 due to its (presumably) higher metal retention efficiency and somewhat different star-formation history. Performing a spatially resolved study of the neutral gas throughout JKB~18 would essentially resolve this question.

As noted, the Leo~P model is not a perfect comparison case for a more massive galaxy like JKB~18, but there is limited theoretical work examining detailed metal abundances in galaxies like JKB~18. While this section aimed to provide insight into possible physical drivers of the observed abundance variations in JKB~18, additional work is required to build a definitive explanation. Examination of the gas-phase abundances of dwarf galaxies in existing and ongoing simulations would be a valuable avenue of future research.

\section{Summary \&\ Conclusions}\label{sec:conc}
The goal of this study was to explore the chemical homogeneity of a low-metallicity, star-forming dwarf galaxy using the high spatial resolution, wide area coverage, IFU data afforded by VLT/MUSE. JKB~18 is a nearby, isolated blue diffuse dwarf galaxy with an average (direct-method) metallicity of $\sim0.08$~\Zsol\ and an average star-formation rate of $0.20\pm0.09\times10^{-2}$~\Msol/yr. We utilized its close proximity to perform a detailed study of its ionized gas in order to explore the distribution of metals throughout its gas and understand the cause and effects of this distribution with the hope of providing insight into the chemical complexity experienced by galaxies in the early Universe.

Emission line maps of \ha, \hb, \foiii, \fnii, \fsii, and \fsiii\, created from the VLT/MUSE data, all revealed JKB~18 to be a system of \hii\ regions without any discernibly ordered structure. However, \ha\ radial velocity maps suggested large-ordered rotation of the gas between $-20$~\kms\ and $+30$~\kms\ relative to the systemic velocity. Little or no signs of gas inflow/outflow from the \hii\ regions were found, with a very uniform velocity dispersion of $\sim$100--120~\kms\ throughout. Emission line diagnostic maps suggest a rather complex ionization structure, with regions of highest ionisation lying offset from the main star-forming regions. Overall, the gas appears to be dominated by photoionization throughout although we cannot rule out shocks due to the low-metallicity of the gas.

Chemical abundance calculations were performed using two methods and the distribution of metals throughout the gas was assessed both statistically and spatially via radial profiles. Firstly, direct method abundances were determined for each of the 30 \hii\ regions throughout JKB~18 by integrating the spectra over the individual regions. Secondly, oxygen abundance maps were derived using the O3N2 strong-line metallicity calibration. For the eleven \hii\ regions for which a direct method abundance was available, abundances were found to have a range of 7.3$\lesssim$12+log(O/H)$\lesssim$7.9, with an average of 12+log(O/H)$=7.6\pm0.2$. Several \hii\ regions were found to lie outside the 1$\sigma$ distribution by $\sim$0.1~dex,  all within 1~kpc of the central (brightest) \hii\ region. A large spread in O/H $\sim0.5$~dex was also found using the high spatial resolution $Z[O3N2]$ map, although without any discernible gradients or global pattern throughout the system. Overall, considering the large size of the 1$\sigma$ distribution, and the fact that several \hii\ regions have O/H values that lie beyond this distribution, we deem JKB~18 to be chemically inhomogeneous. 

In an attempt to understand the cause of this inhomogeneity, we first reviewed similarly irregular chemical abundance distributions found within other dwarf galaxies and the mechanisms causing these anomalies. No relation was found between SFR and low O/H in JKB~18, suggesting that the accretion of metal poor gas is not a dominant factor in influencing the metal content of these regions. In addition to this, no correlation was found between the average age of an \hii\ region's stellar population (found to be between $<5$~Myr and $<$5~Gyr)  and O/H, suggesting short mixing timescales are also not responsible for regions of inhomogeneity.

Secondly, we assessed a high-resolution hydrodynamical galaxy-scale simulation of a low-mass XMP dwarf galaxy. This particular simulation offers possible insight into the chemical inhomogeneity found within JKB~18 in that it allows for `realistic' chemical evolution scenarios. From this simulation we learn that the warm ionized phase of this type of system can indeed be chemically inhomogeneous, with chemical abundance levels largely dependent on the SNe rate. As such, the level of chemical inhomogeneity in dwarf galaxies may be contingent upon both the level of star-formation activity and the timing of our observations with respect to that activity. Overall, the magnitude of inhomogeneity in the model is larger than that found in JKB~18, which we attribute to JKB~18 being more massive than the model and (most likely) having a higher metal retention rate, and being more metal rich and thus effectively `diluting' the levels of metal enrichment experienced from a given event. Additionally, both stellar feedback and environmental effects are thought to play a role in the removal/redistribution of metals from these systems.

Finally, this study not only highlights the fact that (contrary to popular belief) dwarf galaxies \textit{can} be chemically inhomogeneous, but also only draws attention to the biases involved in \textit{determining} whether or not gas is chemically homogeneous. Variations beyond the 1$\sigma$ distribution are not conducive to this decision given that the size of the distribution itself is a measure of chemical inhomogeneity. Moreover, the physical scale of such studies can also play a large role in this measurement, in that sampling the \textit{entire} galaxy on at least \hii\ regions scales ($\sim$100--200~pc) is the only way to provide the detailed, global picture that is required. With the continued usage of large-scale IFUs such as VLT/MUSE and Keck Cosmic Web Imager, large-scale high spatial-resolution studies of galaxy chemical homogeneity will hopefully be more prominent in the literature in the future, allowing us to draw firmer conclusions on the cause and effect of metal variations throughout star-forming galaxies. Moreover, future spatially-resolved spectroscopic studies of the rest-frame UV (e.g., using HST/COS for the nearby Universe or optical IFUs for high-redshift systems) will provide us with much-needed insight into the metal distribution of the cold neutral ISM and a more holisitic understanding of the metal recycling processes ongoing in these systems.

\section*{Acknowledgments}
This study is based on observations collected at the European Organisation for Astronomical Research in the Southern hemisphere under ESO programme(s) 096.B-0212(A). We are grateful to the European Southern Observatory time assignment committee who awarded time to this programme and to the staff astronomers at Paranal who conducted the observations. We thank Peter Zeidler for his invaluable assistance in the MUSE data reduction and for allowing us access to \textsc{MUSEPack} during its development phase, and Roger Wesson for his advice on optimizing ALFA for our line-fitting needs.  We are sincerely grateful to Evan Skillman for discussions concerning chemical inhomogeneities, to Danielle Berg for the assessment of anomalously low metallicity measurements, and Matthew Auger for his assistance with several of the figures presented here. The authors would also like thank the anonymous referee, whose insightful comments helped greatly improve this manuscript. BLJ thanks support from the European Space Agency (ESA). NK thanks the Schlumberger Foundation for the fellowship supporting her postdoctoral work at the University of Cambridge. SK is partially supported by NSF grants AST-1813881, AST-1909584 and Heising-Simon's foundation grant 2018-1030.

\bibliographystyle{mnras}
\bibliography{references}

\begin{thebibliography}{}
\makeatletter
\relax
\def\mn@urlcharsother{\let\do\@makeother \do\$\do\&\do\#\do\^\do\_\do\%\do\~}
\def\mn@doi{\begingroup\mn@urlcharsother \@ifnextchar [ {\mn@doi@}
  {\mn@doi@[]}}
\def\mn@doi@[#1]#2{\def\@tempa{#1}\ifx\@tempa\@empty \href
  {http://dx.doi.org/#2} {doi:#2}\else \href {http://dx.doi.org/#2} {#1}\fi
  \endgroup}
\def\mn@eprint#1#2{\mn@eprint@#1:#2::\@nil}
\def\mn@eprint@arXiv#1{\href {http://arxiv.org/abs/#1} {{\tt arXiv:#1}}}
\def\mn@eprint@dblp#1{\href {http://dblp.uni-trier.de/rec/bibtex/#1.xml}
  {dblp:#1}}
\def\mn@eprint@#1:#2:#3:#4\@nil{\def\@tempa {#1}\def\@tempb {#2}\def\@tempc
  {#3}\ifx \@tempc \@empty \let \@tempc \@tempb \let \@tempb \@tempa \fi \ifx
  \@tempb \@empty \def\@tempb {arXiv}\fi \@ifundefined
  {mn@eprint@\@tempb}{\@tempb:\@tempc}{\expandafter \expandafter \csname
  mn@eprint@\@tempb\endcsname \expandafter{\@tempc}}}

\bibitem[\protect\citeauthoryear{{Annibali}, {Tosi}, {Pasquali}, {Aloisi},
  {Mignoli}  \& {Romano}}{{Annibali} et~al.}{2015}]{Annibali:2015}
{Annibali} F.,  {Tosi} M.,  {Pasquali} A.,  {Aloisi} A.,  {Mignoli} M.,
  {Romano} D.,  2015, \mn@doi [\aj] {10.1088/0004-6256/150/5/143}, \href
  {https://ui.adsabs.harvard.edu/abs/2015AJ....150..143A} {150, 143}

\bibitem[\protect\citeauthoryear{{Annibali} et~al.,}{{Annibali}
  et~al.}{2017}]{Annibali:2017}
{Annibali} F.,  et~al., 2017, \mn@doi [\apj] {10.3847/1538-4357/aa7678}, \href
  {https://ui.adsabs.harvard.edu/abs/2017ApJ...843...20A} {843, 20}

\bibitem[\protect\citeauthoryear{{Annibali} et~al.,}{{Annibali}
  et~al.}{2019}]{Annibali:2019}
{Annibali} F.,  et~al., 2019, \mn@doi [\mnras] {10.1093/mnras/sty2911}, \href
  {https://ui.adsabs.harvard.edu/abs/2019MNRAS.482.3892A} {482, 3892}

\bibitem[\protect\citeauthoryear{{Asplund}, {Grevesse}, {Sauval}  \&
  {Scott}}{{Asplund} et~al.}{2009}]{Asplund:2009}
{Asplund} M.,  {Grevesse} N.,  {Sauval} A.~J.,   {Scott} P.,  2009, \mn@doi
  [\araa] {10.1146/annurev.astro.46.060407.145222}, \href
  {http://adsabs.harvard.edu/abs/2009ARA%26A..47..481A} {47, 481}

\bibitem[\protect\citeauthoryear{{Belfiore} et~al.,}{{Belfiore}
  et~al.}{2017}]{Belfiore:2017}
{Belfiore} F.,  et~al., 2017, \mn@doi [\mnras] {10.1093/mnras/stx789}, \href
  {https://ui.adsabs.harvard.edu/abs/2017MNRAS.469..151B} {469, 151}

\bibitem[\protect\citeauthoryear{{Berg} et~al.,}{{Berg}
  et~al.}{2012}]{Berg:2012}
{Berg} D.~A.,  et~al., 2012, \mn@doi [\apj] {10.1088/0004-637X/754/2/98}, \href
  {http://adsabs.harvard.edu/abs/2012ApJ...754...98B} {754, 98}

\bibitem[\protect\citeauthoryear{{Berg}, {Skillman}, {Croxall}, {Pogge},
  {Moustakas}  \& {Johnson-Groh}}{{Berg} et~al.}{2015}]{Berg:2015}
{Berg} D.~A.,  {Skillman} E.~D.,  {Croxall} K.~V.,  {Pogge} R.~W.,  {Moustakas}
  J.,   {Johnson-Groh} M.,  2015, \mn@doi [\apj] {10.1088/0004-637X/806/1/16},
  \href {http://adsabs.harvard.edu/abs/2015ApJ...806...16B} {806, 16}

\bibitem[\protect\citeauthoryear{{Bresolin}}{{Bresolin}}{2019}]{Bresolin:2019}
{Bresolin} F.,  2019, \mn@doi [\mnras] {10.1093/mnras/stz1947}, \href
  {https://ui.adsabs.harvard.edu/abs/2019MNRAS.488.3826B} {488, 3826}

\bibitem[\protect\citeauthoryear{{Brook}, {Stinson}, {Gibson}, {Shen},
  {Macci{\`o}}, {Obreja}, {Wadsley}  \& {Quinn}}{{Brook}
  et~al.}{2014}]{Brook2014}
{Brook} C.~B.,  {Stinson} G.,  {Gibson} B.~K.,  {Shen} S.,  {Macci{\`o}} A.~V.,
   {Obreja} A.,  {Wadsley} J.,   {Quinn} T.,  2014, \mn@doi [\mnras]
  {10.1093/mnras/stu1406}, \href
  {http://adsabs.harvard.edu/abs/2014MNRAS.443.3809B} {443, 3809}

\bibitem[\protect\citeauthoryear{{Bruzual} \& {Charlot}}{{Bruzual} \&
  {Charlot}}{2003}]{Bruzual:2003}
{Bruzual} G.,  {Charlot} S.,  2003, \mn@doi [\mnras]
  {10.1046/j.1365-8711.2003.06897.x}, \href
  {http://adsabs.harvard.edu/abs/2003MNRAS.344.1000B} {344, 1000}

\bibitem[\protect\citeauthoryear{{Bundy} et~al.,}{{Bundy}
  et~al.}{2015}]{Bundy:2015}
{Bundy} K.,  et~al., 2015, \mn@doi [\apj] {10.1088/0004-637X/798/1/7}, \href
  {http://adsabs.harvard.edu/abs/2015ApJ...798....7B} {798, 7}

\bibitem[\protect\citeauthoryear{{Cid Fernandes}, {Mateus}, {Sodr{\'e}},
  {Stasi{\'n}ska}  \& {Gomes}}{{Cid Fernandes}
  et~al.}{2005}]{CidFernandes:2005}
{Cid Fernandes} R.,  {Mateus} A.,  {Sodr{\'e}} L.,  {Stasi{\'n}ska} G.,
  {Gomes} J.~M.,  2005, \mn@doi [\mnras] {10.1111/j.1365-2966.2005.08752.x},
  \href {https://ui.adsabs.harvard.edu/abs/2005MNRAS.358..363C} {358, 363}

\bibitem[\protect\citeauthoryear{{Corlies}, {Johnston}  \& {Wise}}{{Corlies}
  et~al.}{2018}]{Corlies:2018}
{Corlies} L.,  {Johnston} K.~V.,   {Wise} J.~H.,  2018, \mn@doi [\mnras]
  {10.1093/mnras/sty064}, \href
  {http://adsabs.harvard.edu/abs/2018MNRAS.475.4868C} {475, 4868}

\bibitem[\protect\citeauthoryear{{C{\^o}t{\'e}}, {Silvia}, {O'Shea}, {Smith}
  \& {Wise}}{{C{\^o}t{\'e}} et~al.}{2018}]{Cote:2018}
{C{\^o}t{\'e}} B.,  {Silvia} D.~W.,  {O'Shea} B.~W.,  {Smith} B.,   {Wise}
  J.~H.,  2018, \mn@doi [\apj] {10.3847/1538-4357/aabe8f}, \href
  {https://ui.adsabs.harvard.edu/abs/2018ApJ...859...67C} {859, 67}

\bibitem[\protect\citeauthoryear{{Cresci}, {Mannucci}, {Maiolino}, {Marconi},
  {Gnerucci}  \& {Magrini}}{{Cresci} et~al.}{2010}]{Cresci:2010}
{Cresci} G.,  {Mannucci} F.,  {Maiolino} R.,  {Marconi} A.,  {Gnerucci} A.,
  {Magrini} L.,  2010, \mn@doi [\nat] {10.1038/nature09451}, \href
  {http://adsabs.harvard.edu/abs/2010Natur.467..811C} {467, 811}

\bibitem[\protect\citeauthoryear{{Croom} et~al.,}{{Croom}
  et~al.}{2012}]{Croom:2012}
{Croom} S.~M.,  et~al., 2012, \mn@doi [\mnras]
  {10.1111/j.1365-2966.2011.20365.x}, \href
  {https://ui.adsabs.harvard.edu/abs/2012MNRAS.421..872C} {421, 872}

\bibitem[\protect\citeauthoryear{{Croxall}, {van Zee}, {Lee}, {Skillman},
  {Lee}, {C{\^o}t{\'e}}, {Kennicutt}  \& {Miller}}{{Croxall}
  et~al.}{2009}]{Croxall:2009}
{Croxall} K.~V.,  {van Zee} L.,  {Lee} H.,  {Skillman} E.~D.,  {Lee} J.~C.,
  {C{\^o}t{\'e}} S.,  {Kennicutt} Robert~C. J.,   {Miller} B.~W.,  2009,
  \mn@doi [\apj] {10.1088/0004-637X/705/1/723}, \href
  {https://ui.adsabs.harvard.edu/abs/2009ApJ...705..723C} {705, 723}

\bibitem[\protect\citeauthoryear{{Curti}, {Cresci}, {Mannucci}, {Marconi},
  {Maiolino}  \& {Esposito}}{{Curti} et~al.}{2017}]{Curti:2017}
{Curti} M.,  {Cresci} G.,  {Mannucci} F.,  {Marconi} A.,  {Maiolino} R.,
  {Esposito} S.,  2017, \mn@doi [\mnras] {10.1093/mnras/stw2766}, \href
  {https://ui.adsabs.harvard.edu/abs/2017MNRAS.465.1384C} {465, 1384}

\bibitem[\protect\citeauthoryear{{Dayal}, {Ferrara}  \& {Dunlop}}{{Dayal}
  et~al.}{2013}]{Dayal:2013}
{Dayal} P.,  {Ferrara} A.,   {Dunlop} J.~S.,  2013, \mn@doi [\mnras]
  {10.1093/mnras/stt083}, \href
  {https://ui.adsabs.harvard.edu/abs/2013MNRAS.430.2891D} {430, 2891}

\bibitem[\protect\citeauthoryear{{Edmunds} \& {Roy}}{{Edmunds} \&
  {Roy}}{1993}]{Edmunds:1993}
{Edmunds} M.~G.,  {Roy} J.-R.,  1993, \mn@doi [\mnras]
  {10.1093/mnras/261.1.L17}, \href
  {https://ui.adsabs.harvard.edu/abs/1993MNRAS.261L..17E} {261, L17}

\bibitem[\protect\citeauthoryear{{Ellison}, {Patton}, {Simard}  \&
  {McConnachie}}{{Ellison} et~al.}{2008}]{Ellison:2008}
{Ellison} S.~L.,  {Patton} D.~R.,  {Simard} L.,   {McConnachie} A.~W.,  2008,
  \mn@doi [\apjl] {10.1086/527296}, \href
  {https://ui.adsabs.harvard.edu/abs/2008ApJ...672L.107E} {672, L107}

\bibitem[\protect\citeauthoryear{{Emerick}, {Bryan}, {Mac Low}, {C{\^o}t{\'e}},
  {Johnston}  \& {O'Shea}}{{Emerick} et~al.}{2018}]{Emerick:2018b}
{Emerick} A.,  {Bryan} G.~L.,  {Mac Low} M.-M.,  {C{\^o}t{\'e}} B.,  {Johnston}
  K.~V.,   {O'Shea} B.~W.,  2018, \mn@doi [\apj] {10.3847/1538-4357/aaec7d},
  \href {https://ui.adsabs.harvard.edu/abs/2018ApJ...869...94E} {869, 94}

\bibitem[\protect\citeauthoryear{{Emerick}, {Bryan}  \& {Mac Low}}{{Emerick}
  et~al.}{2019a}]{Emerick:2019b}
{Emerick} A.,  {Bryan} G.~L.,   {Mac Low} M.-M.,  2019a, arXiv e-prints, \href
  {https://ui.adsabs.harvard.edu/abs/2019arXiv190904695E} {p. arXiv:1909.04695}

\bibitem[\protect\citeauthoryear{{Emerick}, {Bryan}  \& {Mac Low}}{{Emerick}
  et~al.}{2019b}]{Emerick:2019}
{Emerick} A.,  {Bryan} G.~L.,   {Mac Low} M.-M.,  2019b, \mn@doi [\mnras]
  {10.1093/mnras/sty2689}, \href
  {https://ui.adsabs.harvard.edu/abs/2019MNRAS.482.1304E} {482, 1304}

\bibitem[\protect\citeauthoryear{{Escala} et~al.,}{{Escala}
  et~al.}{2018}]{Escala2018}
{Escala} I.,  et~al., 2018, \mn@doi [\mnras] {10.1093/mnras/stx2858}, \href
  {http://adsabs.harvard.edu/abs/2018MNRAS.474.2194E} {474, 2194}

\bibitem[\protect\citeauthoryear{{Few}, {Courty}, {Gibson}, {Kawata}, {Calura}
  \& {Teyssier}}{{Few} et~al.}{2012}]{Few:2012}
{Few} C.~G.,  {Courty} S.,  {Gibson} B.~K.,  {Kawata} D.,  {Calura} F.,
  {Teyssier} R.,  2012, \mn@doi [\mnras] {10.1111/j.1745-3933.2012.01275.x},
  \href {http://adsabs.harvard.edu/abs/2012MNRAS.424L..11F} {424, L11}

\bibitem[\protect\citeauthoryear{{Filho}, {S{\'a}nchez Almeida},
  {Mu{\~n}oz-Tu{\~n}{\'o}n}, {Nuza}, {Kitaura}  \& {He{\ss}}}{{Filho}
  et~al.}{2015}]{Filho:2015}
{Filho} M.~E.,  {S{\'a}nchez Almeida} J.,  {Mu{\~n}oz-Tu{\~n}{\'o}n} C.,
  {Nuza} S.~E.,  {Kitaura} F.,   {He{\ss}} S.,  2015, \mn@doi [\apj]
  {10.1088/0004-637X/802/2/82}, \href
  {http://adsabs.harvard.edu/abs/2015ApJ...802...82F} {802, 82}

\bibitem[\protect\citeauthoryear{{Fitzpatrick}}{{Fitzpatrick}}{1999}]{Fitzpatrick:1999}
{Fitzpatrick} E.~L.,  1999, \mn@doi [\pasp] {10.1086/316293}, \href
  {http://adsabs.harvard.edu/abs/1999PASP..111...63F} {111, 63}

\bibitem[\protect\citeauthoryear{{F{\"o}rster Schreiber} et~al.,}{{F{\"o}rster
  Schreiber} et~al.}{2018}]{ForsterSchreiber:2018}
{F{\"o}rster Schreiber} N.~M.,  et~al., 2018, \mn@doi [\apjs]
  {10.3847/1538-4365/aadd49}, \href
  {https://ui.adsabs.harvard.edu/abs/2018ApJS..238...21F} {238, 21}

\bibitem[\protect\citeauthoryear{{Galbany} et~al.,}{{Galbany}
  et~al.}{2016}]{Galbany:2016}
{Galbany} L.,  et~al., 2016, \mn@doi [\mnras] {10.1093/mnras/stv2620}, \href
  {https://ui.adsabs.harvard.edu/abs/2016MNRAS.455.4087G} {455, 4087}

\bibitem[\protect\citeauthoryear{{Giovanelli} et~al.,}{{Giovanelli}
  et~al.}{2013}]{Giovanelli:2013}
{Giovanelli} R.,  et~al., 2013, \mn@doi [\aj] {10.1088/0004-6256/146/1/15},
  \href {http://adsabs.harvard.edu/abs/2013AJ....146...15G} {146, 15}

\bibitem[\protect\citeauthoryear{{Goddard} et~al.,}{{Goddard}
  et~al.}{2017}]{Goddard:2017}
{Goddard} D.,  et~al., 2017, \mn@doi [\mnras] {10.1093/mnras/stw3371}, \href
  {https://ui.adsabs.harvard.edu/abs/2017MNRAS.466.4731G} {466, 4731}

\bibitem[\protect\citeauthoryear{{Hirai} \& {Saitoh}}{{Hirai} \&
  {Saitoh}}{2017}]{Hirai2017}
{Hirai} Y.,  {Saitoh} T.~R.,  2017, \mn@doi [\apjl] {10.3847/2041-8213/aa6799},
  \href {http://adsabs.harvard.edu/abs/2017ApJ...838L..23H} {838, L23}

\bibitem[\protect\citeauthoryear{{James}, {Tsamis}, {Barlow}, {Westmoquette},
  {Walsh}, {Cuisinier}  \& {Exter}}{{James} et~al.}{2009}]{James:2009}
{James} B.~L.,  {Tsamis} Y.~G.,  {Barlow} M.~J.,  {Westmoquette} M.~S.,
  {Walsh} J.~R.,  {Cuisinier} F.,   {Exter} K.~M.,  2009, \mn@doi [\mnras]
  {10.1111/j.1365-2966.2009.15172.x}, \href
  {http://adsabs.harvard.edu/abs/2009MNRAS.398....2J} {398, 2}

\bibitem[\protect\citeauthoryear{{James}, {Tsamis}  \& {Barlow}}{{James}
  et~al.}{2010}]{James:2010}
{James} B.~L.,  {Tsamis} Y.~G.,   {Barlow} M.~J.,  2010, \mn@doi [\mnras]
  {10.1111/j.1365-2966.2009.15706.x}, \href
  {http://adsabs.harvard.edu/abs/2010MNRAS.401..759J} {401, 759}

\bibitem[\protect\citeauthoryear{{James}, {Tsamis}, {Barlow}, {Walsh}  \&
  {Westmoquette}}{{James} et~al.}{2013a}]{James:2013a}
{James} B.~L.,  {Tsamis} Y.~G.,  {Barlow} M.~J.,  {Walsh} J.~R.,
  {Westmoquette} M.~S.,  2013a, \mn@doi [\mnras] {10.1093/mnras/sts004}, \href
  {http://esoads.eso.org/abs/2013MNRAS.428...86J} {428, 86}

\bibitem[\protect\citeauthoryear{{James}, {Tsamis}, {Walsh}, {Barlow}  \&
  {Westmoquette}}{{James} et~al.}{2013b}]{James:2013b}
{James} B.~L.,  {Tsamis} Y.~G.,  {Walsh} J.~R.,  {Barlow} M.~J.,
  {Westmoquette} M.~S.,  2013b, \mn@doi [\mnras] {10.1093/mnras/stt034}, \href
  {http://adsabs.harvard.edu/abs/2013MNRAS.430.2097J} {430, 2097}

\bibitem[\protect\citeauthoryear{James, Auger, Aloisi, Calzetti  \&
  Kewley}{James et~al.}{2015}]{James:2015}
James B.~L.,  Auger M.,  Aloisi A.,  Calzetti D.,   Kewley L.,  2015, \mn@doi
  [The Astrophysical Journal] {10.3847/0004-637x/816/1/40}, 816, 40

\bibitem[\protect\citeauthoryear{James, Koposov, Stark, Belokurov, Pettini,
  Olszewski  \& McQuinn}{James et~al.}{2016a}]{James:2017}
James B.~L.,  Koposov S.~E.,  Stark D.~P.,  Belokurov V.,  Pettini M.,
  Olszewski E.~W.,   McQuinn K. B.~W.,  2016a, \mn@doi [Monthly Notices of the
  Royal Astronomical Society] {10.1093/mnras/stw2962}, 465, 3977

\bibitem[\protect\citeauthoryear{{James}, {Auger}, {Aloisi}, {Calzetti}  \&
  {Kewley}}{{James} et~al.}{2016b}]{James:2016}
{James} B.~L.,  {Auger} M.,  {Aloisi} A.,  {Calzetti} D.,   {Kewley} L.,
  2016b, \mn@doi [\apj] {10.3847/0004-637X/816/1/40}, \href
  {https://ui.adsabs.harvard.edu/abs/2016ApJ...816...40J} {816, 40}

\bibitem[\protect\citeauthoryear{{Jeon}, {Besla}  \& {Bromm}}{{Jeon}
  et~al.}{2017}]{Jeon2017}
{Jeon} M.,  {Besla} G.,   {Bromm} V.,  2017, \mn@doi [\apj]
  {10.3847/1538-4357/aa8c80}, \href
  {http://adsabs.harvard.edu/abs/2017ApJ...848...85J} {848, 85}

\bibitem[\protect\citeauthoryear{{Kehrig}, {V{\'{\i}}lchez}, {S{\'a}nchez},
  {Telles}, {P{\'e}rez-Montero}  \& {Mart{\'{\i}}n-Gord{\'o}n}}{{Kehrig}
  et~al.}{2008}]{Kehrig:2008}
{Kehrig} C.,  {V{\'{\i}}lchez} J.~M.,  {S{\'a}nchez} S.~F.,  {Telles} E.,
  {P{\'e}rez-Montero} E.,   {Mart{\'{\i}}n-Gord{\'o}n} D.,  2008, \mn@doi
  [\aap] {10.1051/0004-6361:20077987}, \href
  {http://adsabs.harvard.edu/abs/2008A%26A...477..813K} {477, 813}

\bibitem[\protect\citeauthoryear{{Kehrig} et~al.,}{{Kehrig}
  et~al.}{2013}]{Kehrig:2013}
{Kehrig} C.,  et~al., 2013, \mn@doi [\mnras] {10.1093/mnras/stt630}, \href
  {http://adsabs.harvard.edu/abs/2013MNRAS.432.2731K} {432, 2731}

\bibitem[\protect\citeauthoryear{{Kennicutt}}{{Kennicutt}}{1998}]{Kennicutt:1998}
{Kennicutt} Jr. R.~C.,  1998, \mn@doi [\araa] {10.1146/annurev.astro.36.1.189},
  \href {http://adsabs.harvard.edu/abs/1998ARA%26A..36..189K} {36, 189}

\bibitem[\protect\citeauthoryear{{Kewley} \& {Ellison}}{{Kewley} \&
  {Ellison}}{2008}]{Kewley:2008}
{Kewley} L.~J.,  {Ellison} S.~L.,  2008, \mn@doi [\apj] {10.1086/587500}, \href
  {http://adsabs.harvard.edu/abs/2008ApJ...681.1183K} {681, 1183}

\bibitem[\protect\citeauthoryear{{Kewley}, {Dopita}, {Sutherland}, {Heisler}
  \& {Trevena}}{{Kewley} et~al.}{2001}]{Kewley:2001}
{Kewley} L.~J.,  {Dopita} M.~A.,  {Sutherland} R.~S.,  {Heisler} C.~A.,
  {Trevena} J.,  2001, \mn@doi [\apj] {10.1086/321545}, \href
  {http://adsabs.harvard.edu/abs/2001ApJ...556..121K} {556, 121}

\bibitem[\protect\citeauthoryear{{Kewley}, {Dopita}, {Leitherer}, {Dav{\'e}},
  {Yuan}, {Allen}, {Groves}  \& {Sutherland}}{{Kewley}
  et~al.}{2013}]{Kewley:2013}
{Kewley} L.~J.,  {Dopita} M.~A.,  {Leitherer} C.,  {Dav{\'e}} R.,  {Yuan} T.,
  {Allen} M.,  {Groves} B.,   {Sutherland} R.,  2013, \mn@doi [\apj]
  {10.1088/0004-637X/774/2/100}, \href
  {http://adsabs.harvard.edu/abs/2013ApJ...774..100K} {774, 100}

\bibitem[\protect\citeauthoryear{{Kobulnicky} \& {Skillman}}{{Kobulnicky} \&
  {Skillman}}{1996}]{Kobulnicky:1996}
{Kobulnicky} H.~A.,  {Skillman} E.~D.,  1996, \mn@doi [\apj] {10.1086/177964},
  \href {http://adsabs.harvard.edu/abs/1996ApJ...471..211K} {471, 211}

\bibitem[\protect\citeauthoryear{{Kobulnicky} \& {Skillman}}{{Kobulnicky} \&
  {Skillman}}{1997}]{Kobulnicky:1997}
{Kobulnicky} H.~A.,  {Skillman} E.~D.,  1997, \apj, \href
  {http://adsabs.harvard.edu/abs/1997ApJ...489..636K} {489, 636}

\bibitem[\protect\citeauthoryear{{Kobulnicky} \& {Skillman}}{{Kobulnicky} \&
  {Skillman}}{1998}]{Kobulnicky:1998}
{Kobulnicky} H.~A.,  {Skillman} E.~D.,  1998, \mn@doi [\apj] {10.1086/305491},
  \href {http://adsabs.harvard.edu/abs/1998ApJ...497..601K} {497, 601}

\bibitem[\protect\citeauthoryear{{Krumholz} \& {Dekel}}{{Krumholz} \&
  {Dekel}}{2012}]{Krumholz:2012}
{Krumholz} M.~R.,  {Dekel} A.,  2012, \mn@doi [\apj]
  {10.1088/0004-637X/753/1/16}, \href
  {https://ui.adsabs.harvard.edu/abs/2012ApJ...753...16K} {753, 16}

\bibitem[\protect\citeauthoryear{{Kudritzki} \& {Puls}}{{Kudritzki} \&
  {Puls}}{2000}]{Kudritzki:2000}
{Kudritzki} R.-P.,  {Puls} J.,  2000, \mn@doi [\araa]
  {10.1146/annurev.astro.38.1.613}, \href
  {https://ui.adsabs.harvard.edu/abs/2000ARA&A..38..613K} {38, 613}

\bibitem[\protect\citeauthoryear{{Kumari}, {James}  \& {Irwin}}{{Kumari}
  et~al.}{2017}]{Kumari:2017}
{Kumari} N.,  {James} B.~L.,   {Irwin} M.~J.,  2017, \mn@doi [\mnras]
  {10.1093/mnras/stx1414}, \href
  {https://ui.adsabs.harvard.edu/abs/2017MNRAS.470.4618K} {470, 4618}

\bibitem[\protect\citeauthoryear{{Kumari}, {James}, {Irwin}, {Amor{\'\i}n}  \&
  {P{\'e}rez-Montero}}{{Kumari} et~al.}{2018}]{Kumari:2018}
{Kumari} N.,  {James} B.~L.,  {Irwin} M.~J.,  {Amor{\'\i}n} R.,
  {P{\'e}rez-Montero} E.,  2018, \mn@doi [\mnras] {10.1093/mnras/sty402}, \href
  {https://ui.adsabs.harvard.edu/abs/2018MNRAS.476.3793K} {476, 3793}

\bibitem[\protect\citeauthoryear{{Kumari}, {Maiolino}, {Belfiore}  \&
  {Curti}}{{Kumari} et~al.}{2019a}]{Kumari:2019b}
{Kumari} N.,  {Maiolino} R.,  {Belfiore} F.,   {Curti} M.,  2019a, \mn@doi
  [\mnras] {10.1093/mnras/stz366}, \href
  {https://ui.adsabs.harvard.edu/abs/2019MNRAS.485..367K} {485, 367}

\bibitem[\protect\citeauthoryear{{Kumari}, {James}, {Irwin}  \&
  {Aloisi}}{{Kumari} et~al.}{2019b}]{Kumari:2019a}
{Kumari} N.,  {James} B.~L.,  {Irwin} M.~J.,   {Aloisi} A.,  2019b, \mn@doi
  [\mnras] {10.1093/mnras/stz343}, \href
  {https://ui.adsabs.harvard.edu/abs/2019MNRAS.485.1103K} {485, 1103}

\bibitem[\protect\citeauthoryear{Lagos, Telles, Nigoche-Netro  \&
  Carrasco}{Lagos et~al.}{2012}]{Lagos:2012}
Lagos P.,  Telles E.,  Nigoche-Netro A.,   Carrasco E.~R.,  2012, \mn@doi
  [Monthly Notices of the Royal Astronomical Society]
  {10.1111/j.1365-2966.2012.21944.x}, 427

\bibitem[\protect\citeauthoryear{{Lagos}, {Papaderos}, {Gomes}, {Smith
  Castelli}  \& {Vega}}{{Lagos} et~al.}{2014}]{Lagos:2014}
{Lagos} P.,  {Papaderos} P.,  {Gomes} J.~M.,  {Smith Castelli} A.~V.,   {Vega}
  L.~R.,  2014, \mn@doi [\aap] {10.1051/0004-6361/201323353}, \href
  {https://ui.adsabs.harvard.edu/abs/2014A&A...569A.110L} {569, A110}

\bibitem[\protect\citeauthoryear{{Lagos}, {Demarco}, {Papaderos}, {Telles},
  {Nigoche-Netro}, {Humphrey}, {Roche}  \& {Gomes}}{{Lagos}
  et~al.}{2016}]{Lagos:2016}
{Lagos} P.,  {Demarco} R.,  {Papaderos} P.,  {Telles} E.,  {Nigoche-Netro} A.,
  {Humphrey} A.,  {Roche} N.,   {Gomes} J.~M.,  2016, \mn@doi [\mnras]
  {10.1093/mnras/stv2702}, \href
  {https://ui.adsabs.harvard.edu/abs/2016MNRAS.456.1549L} {456, 1549}

\bibitem[\protect\citeauthoryear{{Lagos}, {Scott}, {Nigoche-Netro}, {Demarco},
  {Humphrey}  \& {Papaderos}}{{Lagos} et~al.}{2018}]{Lagos:2018}
{Lagos} P.,  {Scott} T.~C.,  {Nigoche-Netro} A.,  {Demarco} R.,  {Humphrey} A.,
    {Papaderos} P.,  2018, \mn@doi [\mnras] {10.1093/mnras/sty601}, \href
  {https://ui.adsabs.harvard.edu/abs/2018MNRAS.477..392L} {477, 392}

\bibitem[\protect\citeauthoryear{{Lee}, {Skillman}  \& {Venn}}{{Lee}
  et~al.}{2006}]{Lee:2006}
{Lee} H.,  {Skillman} E.~D.,   {Venn} K.~A.,  2006, \mn@doi [\apj]
  {10.1086/500568}, \href
  {https://ui.adsabs.harvard.edu/abs/2006ApJ...642..813L} {642, 813}

\bibitem[\protect\citeauthoryear{{Lee} et~al.,}{{Lee} et~al.}{2009}]{Lee:2009}
{Lee} J.~C.,  et~al., 2009, \mn@doi [\apj] {10.1088/0004-637X/706/1/599}, \href
  {http://adsabs.harvard.edu/abs/2009ApJ...706..599L} {706, 599}

\bibitem[\protect\citeauthoryear{{Lehnert}, {van Driel}, {Le Tiran}, {Di
  Matteo}  \& {Haywood}}{{Lehnert} et~al.}{2015}]{Lehnert:2015}
{Lehnert} M.~D.,  {van Driel} W.,  {Le Tiran} L.,  {Di Matteo} P.,   {Haywood}
  M.,  2015, \mn@doi [\aap] {10.1051/0004-6361/201322630}, \href
  {https://ui.adsabs.harvard.edu/abs/2015A&A...577A.112L} {577, A112}

\bibitem[\protect\citeauthoryear{{Lequeux}, {Peimbert}, {Rayo}, {Serrano}  \&
  {Torres-Peimbert}}{{Lequeux} et~al.}{1979}]{Lequeux:1979}
{Lequeux} J.,  {Peimbert} M.,  {Rayo} J.~F.,  {Serrano} A.,   {Torres-Peimbert}
  S.,  1979, \aap, \href {http://adsabs.harvard.edu/abs/1979A%26A....80..155L}
  {80, 155}

\bibitem[\protect\citeauthoryear{{Lilly}, {Carollo}, {Pipino}, {Renzini}  \&
  {Peng}}{{Lilly} et~al.}{2013}]{Lilly:2013}
{Lilly} S.~J.,  {Carollo} C.~M.,  {Pipino} A.,  {Renzini} A.,   {Peng} Y.,
  2013, \mn@doi [\apj] {10.1088/0004-637X/772/2/119}, \href
  {http://adsabs.harvard.edu/abs/2013ApJ...772..119L} {772, 119}

\bibitem[\protect\citeauthoryear{{L{\'o}pez-S{\'a}nchez}, {Dopita}, {Kewley},
  {Zahid}, {Nicholls}  \& {Scharw{\"a}chter}}{{L{\'o}pez-S{\'a}nchez}
  et~al.}{2012}]{Lopez-Sanchez:2012}
{L{\'o}pez-S{\'a}nchez} {\'A}.~R.,  {Dopita} M.~A.,  {Kewley} L.~J.,  {Zahid}
  H.~J.,  {Nicholls} D.~C.,   {Scharw{\"a}chter} J.,  2012, \mn@doi [\mnras]
  {10.1111/j.1365-2966.2012.21145.x}, \href
  {http://adsabs.harvard.edu/abs/2012MNRAS.426.2630L} {426, 2630}

\bibitem[\protect\citeauthoryear{Maiolino \& Mannucci}{Maiolino \&
  Mannucci}{2019}]{Maiolino:2019}
Maiolino R.,  Mannucci F.,  2019, \mn@doi [The Astronomy and Astrophysics
  Review] {10.1007/s00159-018-0112-2}, 27, 3

\bibitem[\protect\citeauthoryear{{Mannucci}, {Cresci}, {Maiolino}, {Marconi}
  \& {Gnerucci}}{{Mannucci} et~al.}{2010}]{Mannucci:2010}
{Mannucci} F.,  {Cresci} G.,  {Maiolino} R.,  {Marconi} A.,   {Gnerucci} A.,
  2010, \mn@doi [\mnras] {10.1111/j.1365-2966.2010.17291.x}, \href
  {http://adsabs.harvard.edu/abs/2010MNRAS.408.2115M} {408, 2115}

\bibitem[\protect\citeauthoryear{{Mast} et~al.,}{{Mast}
  et~al.}{2014}]{Mast:2014}
{Mast} D.,  et~al., 2014, \mn@doi [\aap] {10.1051/0004-6361/201321789}, \href
  {http://adsabs.harvard.edu/abs/2014A%26A...561A.129M} {561, A129}

\bibitem[\protect\citeauthoryear{{McQuinn} et~al.,}{{McQuinn}
  et~al.}{2015}]{McQuinn:2015a}
{McQuinn} K.~B.~W.,  et~al., 2015, \mn@doi [\apj]
  {10.1088/0004-637X/812/2/158}, \href
  {http://adsabs.harvard.edu/abs/2015ApJ...812..158M} {812, 158}

\bibitem[\protect\citeauthoryear{{Moustakas}, {Kennicutt}, {Tremonti}, {Dale},
  {Smith}  \& {Calzetti}}{{Moustakas} et~al.}{2010}]{Moustakas:2010}
{Moustakas} J.,  {Kennicutt} Robert~C. J.,  {Tremonti} C.~A.,  {Dale} D.~A.,
  {Smith} J.-D.~T.,   {Calzetti} D.,  2010, \mn@doi [\apjs]
  {10.1088/0067-0049/190/2/233}, \href
  {https://ui.adsabs.harvard.edu/abs/2010ApJS..190..233M} {190, 233}

\bibitem[\protect\citeauthoryear{{Pagel} \& {Edmunds}}{{Pagel} \&
  {Edmunds}}{1981}]{Pagel:1981}
{Pagel} B.~E.~J.,  {Edmunds} M.~G.,  1981, \mn@doi [\araa]
  {10.1146/annurev.aa.19.090181.000453}, \href
  {https://ui.adsabs.harvard.edu/abs/1981ARA&A..19...77P} {19, 77}

\bibitem[\protect\citeauthoryear{{Papaderos}, {Guseva}, {Izotov}  \&
  {Fricke}}{{Papaderos} et~al.}{2008}]{Papaderos:2008}
{Papaderos} P.,  {Guseva} N.~G.,  {Izotov} Y.~I.,   {Fricke} K.~J.,  2008,
  \mn@doi [\aap] {10.1051/0004-6361:200810028}, \href
  {http://adsabs.harvard.edu/abs/2008A%26A...491..113P} {491, 113}

\bibitem[\protect\citeauthoryear{{Peng} \& {Maiolino}}{{Peng} \&
  {Maiolino}}{2014}]{Peng:2014}
{Peng} Y.-j.,  {Maiolino} R.,  2014, \mn@doi [\mnras] {10.1093/mnras/stu1288},
  \href {https://ui.adsabs.harvard.edu/abs/2014MNRAS.443.3643P} {443, 3643}

\bibitem[\protect\citeauthoryear{{P{\'e}rez-Montero}
  et~al.,}{{P{\'e}rez-Montero} et~al.}{2011}]{Perez-Montero:2011}
{P{\'e}rez-Montero} E.,  et~al., 2011, \mn@doi [\aap]
  {10.1051/0004-6361/201116582}, \href
  {http://adsabs.harvard.edu/abs/2011A%26A...532A.141P} {532, A141}

\bibitem[\protect\citeauthoryear{{Pettini} \& {Pagel}}{{Pettini} \&
  {Pagel}}{2004}]{Pettini:2004}
{Pettini} M.,  {Pagel} B.~E.~J.,  2004, \mn@doi [\mnras]
  {10.1111/j.1365-2966.2004.07591.x}, \href
  {http://adsabs.harvard.edu/abs/2004MNRAS.348L..59P} {348, L59}

\bibitem[\protect\citeauthoryear{{Plat}, {Charlot}, {Bruzual}, {Feltre},
  {Vidal-Garc{\'\i}a}, {Morisset}, {Chevallard}  \& {Todt}}{{Plat}
  et~al.}{2019}]{Plat:2019}
{Plat} A.,  {Charlot} S.,  {Bruzual} G.,  {Feltre} A.,  {Vidal-Garc{\'\i}a} A.,
   {Morisset} C.,  {Chevallard} J.,   {Todt} H.,  2019, \mn@doi [\mnras]
  {10.1093/mnras/stz2616}, \href
  {https://ui.adsabs.harvard.edu/abs/2019MNRAS.tmp.2242P} {p.~2242}

\bibitem[\protect\citeauthoryear{{Poetrodjojo} et~al.,}{{Poetrodjojo}
  et~al.}{2018}]{Poetrodjojo:2018}
{Poetrodjojo} H.,  et~al., 2018, \mn@doi [\mnras] {10.1093/mnras/sty1782},
  \href {https://ui.adsabs.harvard.edu/abs/2018MNRAS.479.5235P} {479, 5235}

\bibitem[\protect\citeauthoryear{{Putman}}{{Putman}}{2017}]{Putman:2017}
{Putman} M.~E.,  2017, in {Fox} A.,  {Dav{\'e}} R.,  eds,  Astrophysics and
  Space Science Library Vol. 430, Gas Accretion onto Galaxies. p.~1 (\mn@eprint
  {arXiv} {1612.00461}), \mn@doi{10.1007/978-3-319-52512-9_1}

\bibitem[\protect\citeauthoryear{{Revaz}, {Arnaudon}, {Nichols}, {Bonvin}  \&
  {Jablonka}}{{Revaz} et~al.}{2016}]{Revaz2016}
{Revaz} Y.,  {Arnaudon} A.,  {Nichols} M.,  {Bonvin} V.,   {Jablonka} P.,
  2016, \mn@doi [\aap] {10.1051/0004-6361/201526438}, \href
  {http://adsabs.harvard.edu/abs/2016A%26A...588A..21R} {588, A21}

\bibitem[\protect\citeauthoryear{{Ritter}, {Herwig}, {Jones}, {Pignatari},
  {Fryer}  \& {Hirschi}}{{Ritter} et~al.}{2018}]{Ritter:2018}
{Ritter} C.,  {Herwig} F.,  {Jones} S.,  {Pignatari} M.,  {Fryer} C.,
  {Hirschi} R.,  2018, \mn@doi [\mnras] {10.1093/mnras/sty1729}, \href
  {http://adsabs.harvard.edu/abs/2018MNRAS.480..538R} {480, 538}

\bibitem[\protect\citeauthoryear{{S{\'a}nchez Almeida}, {Morales-Luis},
  {Mu{\~n}oz-Tu{\~n}{\'o}n}, {Elmegreen}, {Elmegreen}  \&
  {M{\'e}ndez-Abreu}}{{S{\'a}nchez Almeida}
  et~al.}{2014}]{Sanchez-Almeida:2014b}
{S{\'a}nchez Almeida} J.,  {Morales-Luis} A.~B.,  {Mu{\~n}oz-Tu{\~n}{\'o}n} C.,
   {Elmegreen} D.~M.,  {Elmegreen} B.~G.,   {M{\'e}ndez-Abreu} J.,  2014,
  \mn@doi [\apj] {10.1088/0004-637X/783/1/45}, \href
  {http://adsabs.harvard.edu/abs/2014ApJ...783...45S} {783, 45}

\bibitem[\protect\citeauthoryear{{S{\'a}nchez Almeida}, {P{\'e}rez-Montero},
  {Morales-Luis}, {Mu{\~n}oz-Tu{\~n}{\'o}n}, {Garc{\'{\i}}a-Benito}, {Nuza}  \&
  {Kitaura}}{{S{\'a}nchez Almeida} et~al.}{2016}]{Sanchez-Almeida:2016}
{S{\'a}nchez Almeida} J.,  {P{\'e}rez-Montero} E.,  {Morales-Luis} A.~B.,
  {Mu{\~n}oz-Tu{\~n}{\'o}n} C.,  {Garc{\'{\i}}a-Benito} R.,  {Nuza} S.~E.,
  {Kitaura} F.~S.,  2016, \mn@doi [\apj] {10.3847/0004-637X/819/2/110}, \href
  {http://adsabs.harvard.edu/abs/2016ApJ...819..110S} {819, 110}

\bibitem[\protect\citeauthoryear{{S{\'a}nchez} et~al.,}{{S{\'a}nchez}
  et~al.}{2012}]{Sanchez:2012}
{S{\'a}nchez} S.~F.,  et~al., 2012, \mn@doi [\aap]
  {10.1051/0004-6361/201219578}, \href
  {https://ui.adsabs.harvard.edu/abs/2012A&A...546A...2S} {546, A2}

\bibitem[\protect\citeauthoryear{{Searle}}{{Searle}}{1971}]{Searle:1971}
{Searle} L.,  1971, \mn@doi [\apj] {10.1086/151090}, \href
  {https://ui.adsabs.harvard.edu/abs/1971ApJ...168..327S} {168, 327}

\bibitem[\protect\citeauthoryear{{Shen}, {Wadsley}  \& {Stinson}}{{Shen}
  et~al.}{2010}]{Shen2010}
{Shen} S.,  {Wadsley} J.,   {Stinson} G.,  2010, \mn@doi [\mnras]
  {10.1111/j.1365-2966.2010.17047.x}, \href
  {http://adsabs.harvard.edu/abs/2010MNRAS.407.1581S} {407, 1581}

\bibitem[\protect\citeauthoryear{{Shen}, {Madau}, {Guedes}, {Mayer},
  {Prochaska}  \& {Wadsley}}{{Shen} et~al.}{2013}]{Shen2013}
{Shen} S.,  {Madau} P.,  {Guedes} J.,  {Mayer} L.,  {Prochaska} J.~X.,
  {Wadsley} J.,  2013, \mn@doi [\apj] {10.1088/0004-637X/765/2/89}, \href
  {http://adsabs.harvard.edu/abs/2013ApJ...765...89S} {765, 89}

\bibitem[\protect\citeauthoryear{{Skillman}, {Kennicutt}  \&
  {Hodge}}{{Skillman} et~al.}{1989}]{Skillman:1989}
{Skillman} E.~D.,  {Kennicutt} R.~C.,   {Hodge} P.~W.,  1989, \mn@doi [\apj]
  {10.1086/168178}, \href {http://adsabs.harvard.edu/abs/1989ApJ...347..875S}
  {347, 875}

\bibitem[\protect\citeauthoryear{{Skillman} et~al.,}{{Skillman}
  et~al.}{2013}]{Skillman:2013}
{Skillman} E.~D.,  et~al., 2013, \mn@doi [\aj] {10.1088/0004-6256/146/1/3},
  \href {http://adsabs.harvard.edu/abs/2013AJ....146....3S} {146, 3}

\bibitem[\protect\citeauthoryear{{Su}, {Hopkins}, {Hayward},
  {Faucher-Gigu{\`e}re}, {Kere{\v s}}, {Ma}  \& {Robles}}{{Su}
  et~al.}{2017}]{Su2017a}
{Su} K.-Y.,  {Hopkins} P.~F.,  {Hayward} C.~C.,  {Faucher-Gigu{\`e}re} C.-A.,
  {Kere{\v s}} D.,  {Ma} X.,   {Robles} V.~H.,  2017, \mn@doi [\mnras]
  {10.1093/mnras/stx1463}, \href
  {http://adsabs.harvard.edu/abs/2017MNRAS.471..144S} {471, 144}

\bibitem[\protect\citeauthoryear{{Tenorio-Tagle}}{{Tenorio-Tagle}}{1996}]{Tenorio-Tagle:1996}
{Tenorio-Tagle} G.,  1996, \mn@doi [\aj] {10.1086/117903}, \href
  {https://ui.adsabs.harvard.edu/abs/1996AJ....111.1641T} {111, 1641}

\bibitem[\protect\citeauthoryear{{Tremonti} et~al.,}{{Tremonti}
  et~al.}{2004}]{Tremonti:2004}
{Tremonti} C.~A.,  et~al., 2004, \mn@doi [\apj] {10.1086/423264}, \href
  {https://ui.adsabs.harvard.edu/abs/2004ApJ...613..898T} {613, 898}

\bibitem[\protect\citeauthoryear{{Troncoso} et~al.,}{{Troncoso}
  et~al.}{2014}]{Troncoso:2014}
{Troncoso} P.,  et~al., 2014, \mn@doi [\aap] {10.1051/0004-6361/201322099},
  \href {https://ui.adsabs.harvard.edu/abs/2014A&A...563A..58T} {563, A58}

\bibitem[\protect\citeauthoryear{{Vila-Costas} \& {Edmunds}}{{Vila-Costas} \&
  {Edmunds}}{1992}]{Vila-Costas:1992}
{Vila-Costas} M.~B.,  {Edmunds} M.~G.,  1992, \mn@doi [\mnras]
  {10.1093/mnras/259.1.121}, \href
  {https://ui.adsabs.harvard.edu/abs/1992MNRAS.259..121V} {259, 121}

\bibitem[\protect\citeauthoryear{{Werk}, {Putman}, {Meurer}  \&
  {Santiago-Figueroa}}{{Werk} et~al.}{2011}]{Werk:2011}
{Werk} J.~K.,  {Putman} M.~E.,  {Meurer} G.~R.,   {Santiago-Figueroa} N.,
  2011, \mn@doi [\apj] {10.1088/0004-637X/735/2/71}, \href
  {https://ui.adsabs.harvard.edu/abs/2011ApJ...735...71W} {735, 71}

\bibitem[\protect\citeauthoryear{{Wesson}}{{Wesson}}{2016}]{Wesson:2016}
{Wesson} R.,  2016, \mn@doi [\mnras] {10.1093/mnras/stv2946}, \href
  {https://ui.adsabs.harvard.edu/abs/2016MNRAS.456.3774W} {456, 3774}

\bibitem[\protect\citeauthoryear{{Westmoquette}, {James}, {Monreal-Ibero}  \&
  {Walsh}}{{Westmoquette} et~al.}{2013}]{Westmoquette:2013}
{Westmoquette} M.~S.,  {James} B.,  {Monreal-Ibero} A.,   {Walsh} J.~R.,  2013,
  \mn@doi [\aap] {10.1051/0004-6361/201220580}, \href
  {http://adsabs.harvard.edu/abs/2013A%26A...550A..88W} {550, A88}

\bibitem[\protect\citeauthoryear{{Yuan}, {Kewley}  \& {Rich}}{{Yuan}
  et~al.}{2013}]{Yuan:2013}
{Yuan} T.-T.,  {Kewley} L.~J.,   {Rich} J.,  2013, \mn@doi [\apj]
  {10.1088/0004-637X/767/2/106}, \href
  {http://adsabs.harvard.edu/abs/2013ApJ...767..106Y} {767, 106}

\bibitem[\protect\citeauthoryear{{Zaritsky}, {Kennicutt}  \&
  {Huchra}}{{Zaritsky} et~al.}{1994}]{Zaritsky:1994}
{Zaritsky} D.,  {Kennicutt} Robert~C. J.,   {Huchra} J.~P.,  1994, \mn@doi
  [\apj] {10.1086/173544}, \href
  {https://ui.adsabs.harvard.edu/abs/1994ApJ...420...87Z} {420, 87}

\bibitem[\protect\citeauthoryear{{Zeidler}, {Nota}, {Sabbi}, {Luljak},
  {McLeod}, {Grebel}, {Pasquali}  \& {Tosi}}{{Zeidler}
  et~al.}{2019}]{Zeidler:2019}
{Zeidler} P.,  {Nota} A.,  {Sabbi} E.,  {Luljak} P.,  {McLeod} A.~F.,  {Grebel}
  E.~K.,  {Pasquali} A.,   {Tosi} M.,  2019, arXiv e-prints, \href
  {https://ui.adsabs.harvard.edu/abs/2019arXiv190908143Z} {p. arXiv:1909.08143}

\bibitem[\protect\citeauthoryear{{van Zee}, {Salzer}, {Haynes}, {O'Donoghue}
  \& {Balonek}}{{van Zee} et~al.}{1998}]{VanZee:1998}
{van Zee} L.,  {Salzer} J.~J.,  {Haynes} M.~P.,  {O'Donoghue} A.~A.,
  {Balonek} T.~J.,  1998, \mn@doi [\aj] {10.1086/300647}, \href
  {https://ui.adsabs.harvard.edu/abs/1998AJ....116.2805V} {116, 2805}

\makeatother
\end{thebibliography}

\end{document}